\documentclass[showkeys, showpacs]{revtex4}
\usepackage{color}
\usepackage{graphicx}

\begin{document}

\title{ Special Relativity for the Full Speed Range\\

---  speed slower than $C_R$ also equal to and faster than $C_R$  }
\author{Youshan Dai$^1$}\email{daiys@zucc.edu.cn}
\author{Kang Li$^2$}\email{kangli@hznu.edu.cn}
\affiliation{1, Zhejiang University City College, Hangzhou, 310015, P.R. China}
\affiliation{2,Department of Physics, Hangzhou Normal University, Hangzhou, 310036, P.R. China}

\date{\today}

\begin{abstract}
 In this paper, we establish a theory of Special Relativity valid for the entire speed range without the assumption of constant speed of light. Two particles species are defined, one species of particles have rest frames with rest mass, and another species of particles do not have rest frame and can not define rest mass. We prove that for the particles which have rest frames, the Galilean transformation is the only linear transformation of space-time that allows infinite speed of particle motion. Hence without any assumption, an upper bound of speed is required for all non-Galilean linear transformations. We then present a novel derivation of the mass-velocity and the mass-energy relations in the framework of relativistic dynamics, which is solely based on the principle of relativity and basic definitions of relativistic momentum and energy. The generalized Lorentz transformation is then determined. The new relativistic formulas are not related directly to the speed of light $c$, but are replaced by a $Relativity~Constant~C_R$ which is an universal speed constant of the Nature introduced in relativistic dynamics. The value of $C_R$ should be measured in experiments, and the usual Lorentz transformation is recovered when setting $C_R=c$. Particles having rest mass and moving slower than $C_R$ are called $tardyons$. Particles having neither rest frames nor rest mass and moving faster than $C_R$ are called $tachyons$, and with the real mass-velocity relation $m=|\vec{p}_\infty |(v^2-C_R^2)^{-\frac{1}{2}}$ where $\vec{p}_\infty $ is the finite momentum of tachyon at infinite speed. Particles with constant-speed $C_R$, also having neither rest frames nor rest mass, are called $constons$. For all particles, $p^2=\vec{p}^2-({E^2}/{C_R^2})$ remains invariant under transformations between inertia frames. The invariant reads $p^2=-m_0^2C_R^2 <0$ for tardyons, $p^2 =0$ for constons and $p^2=|\vec{p}_\infty|^2 >0$ for tachyons, respectively. Thus a new Special Relativity is developed which can be applied both to particles having rest frames with $|\vec{v}| < C_R$ and particles having no rest frames with $|\vec{v}| \geq C_R$.
\end{abstract}


\maketitle


\section{Introduction}

It has been well known that Einstein formulated the Special Relativity (relativity) based on two fundamental postulations: light has constant speed in vacuum (constant speed of light), and the Einstein's principle of relativity~\cite{1,2,3,4}. From the postulation of constant speed of light, the Lorentz transformation of space-time is derived, which lays the very foundation of relativistic kinematics~\cite{5}. The Einstein principle of relativity dictates that all physics laws take covariant forms in all inertial frames, among which no particular frame is physically special. The relativistic mass-velocity relation $m=m_0[1-(v^2/c^2)]^{-\frac{1}{2}}$ and Einstein's mass-energy equation  $E=mc^2$ then follows naturally from those postulations~\cite{6,7}. The relativistic mass-velocity relation and mass-energy relation are basic equations of relativistic dynamics. Traditional derivations of them utilize the Lorentz transformation, tempting one to conclude that they have to be established from the Lorentz transformation rule of the space-time. The Lorentz transformation for velocity, parameterized by the speed of light, implies that velocity of all particles (objects) cannot exceed a speed limit, which is commonly appreciated as a consequence of constant speed of light. It has been proposed that instead of assuming constant speed of light, one can assume the existence of an upper bound for speed. Still, one implicitly assumes that there is some particle species that travel at the speed limit (e.g. the photon), from which the Lorentz transformation formulations rely on~\cite{8,9,10}. In usual formulations of Special Relativity, the relativistic energy-momentum relation is $E^2-\vec{p}^2c^2 =m_0^2c^4 >0$. Particles with velocity less than speed of light have non-zero, real rest mass, and the photon should have zero rest mass respectively, while superluminal particles must have imaginary rest mass~\cite{11,12}. However, the photon and possibly the superluminal particles actually have no rest frames under any Lorentz transformations, so we believe that the notion of rest mass for those without rest frames is meaningless.

In this paper, we take a novel route and formulate a new form of Special Relativity valid for the entire speed range. A $Relativity~Constant~C_R$ will be introduced as required by the principle of relativity, and the assumption of constant speed of light will be shown unnecessary. Particles with rest frames and rest mass, the $tardyons$, have upper speed limit $0\leq |\vec{v}|<C_R$. Particles having neither rest frames nor rest mass, the $ tachyons$, are subject to lower speed limit $C_R<|\vec{v}|<\infty$. In spite that the rest mass can not be defined for tachyons, they have intrinsic momentum and real mass at $|\vec{v}|>C_R$. Finally, particles that travel always at constant-speed $|\vec{v}|=C_R$ in all inertial frames, called $constons$, also have no rest frame and lack a definition of the rest mass. In the case that the Relativity Constant is equal to the speed of light $C_R=c$, the photon will be a conston. Our work shows that the Special Relativity holds for the full speed range consistently, regardless of whether or not particles having no rest frame and with speed $|\vec{v}| \geq C_R$ exist.

The paper is organized as follows: In Section~\ref{sec: kinematics}, we formulate the kinematics of the Special Relativity without the assumption of constant speed of light. We prove that for particles having rest frames, the Galilean relativity with the notion of absolute time is the only one that accommodates a linear transformation of space-time and at the same time allows infinite particle velocity. It implies that for any non-Galilean relativity, upper speed limit for massive particles arises naturally without additional assumptions or knowledge of space-time. Although by definition, particle's velocity can be arbitrarily close to the speed limit, it is not {\it a priori} that any particular species saturates the bound. Therefore without other assumptions, the explicit form of non-Galilean transformations of space-time is not determined in relativistic kinematics. In Section~\ref{sec:dynamics}, for massive particles with $|\vec{v}|<C_R$, the dynamics of relativity is formulated through a new derivation of the mass-velocity and mass-energy relations which does not rely on any explicit space-time transforming rule and which is solely based on the principle of relativity as well as basic dynamic definitions $\vec{p}= m\vec{v}$, $d\vec{p}= \vec{F}dt$, $dE=\vec{F}\cdot d\vec{r}$. The generalized Lorentz transformation for space-time is then determined without the assumption of the constant speed of light. For particle species without rest frames, we show that $C_R$ serves as a lower speed bound for tachyons with $|\vec{v}|> C_R$, and the tachyonic mass must be a real parameter. In Section~\ref{sec: energy-momentum-transform} we study the relativistic transformation for energy and momentum. For all particles, the momentum and energy read $\vec{p}= m\vec{v}$, $E= mC_R^2$ respectively. The energy-momentum relation for either tardyons or tachyons can be obtained by eliminating $m$ from the corresponding mass-velocity relation. The quantity $E^2-\vec{p}^2C_R^2 $ is then shown to be invariant under boosts between inertial frames, and again the generalized Lorentz transformation can then be determined uniquely without the assumption of constant speed of light. Concluding remarks are given in Section~\ref{sec:conclusion}, where we claim that our new formulation of Special Relativity is more general than the traditional formulation, which applies to both particles having rest frames with $|\vec{v}|<C_R$ and particles having no rest frames with $|\vec{v}| \geq C_R$.


\section{Special relativistic kinematics}
\label{sec: kinematics}

The classical Newtonian kinematics are based on the notion of absolute time $t'=t$ and lead to the Galilean transformation. Likewise, the traditional relativistic kinematics are based on the assumption of constant speed of light and lead to the Lorentz transformation. In this Section, we discuss the properties of general linear transformation of space-time without introducing any assumption. For the particles without rest frames, i.e. $\vec{v}\neq 0$ in any case, obviously there is a non-zero lower speed limit $v_{min}$ for them: $0<v_{min}<|\vec{v}|< \infty$. But for the particles which have rest frames, we prove that the Galilean transformation is the only linear space-time transformation that allows infinite speed for particle motion, it implies that for any non-Galilean linear transformation, an upper bound for speed $v_{max}$ is required: $0\leq |\vec{v}|<v_{max}$ without any additional assumption.

\subsection{\bf Linear space-time transformation}

\begin{figure}[ht]
  \includegraphics[scale=0.4]{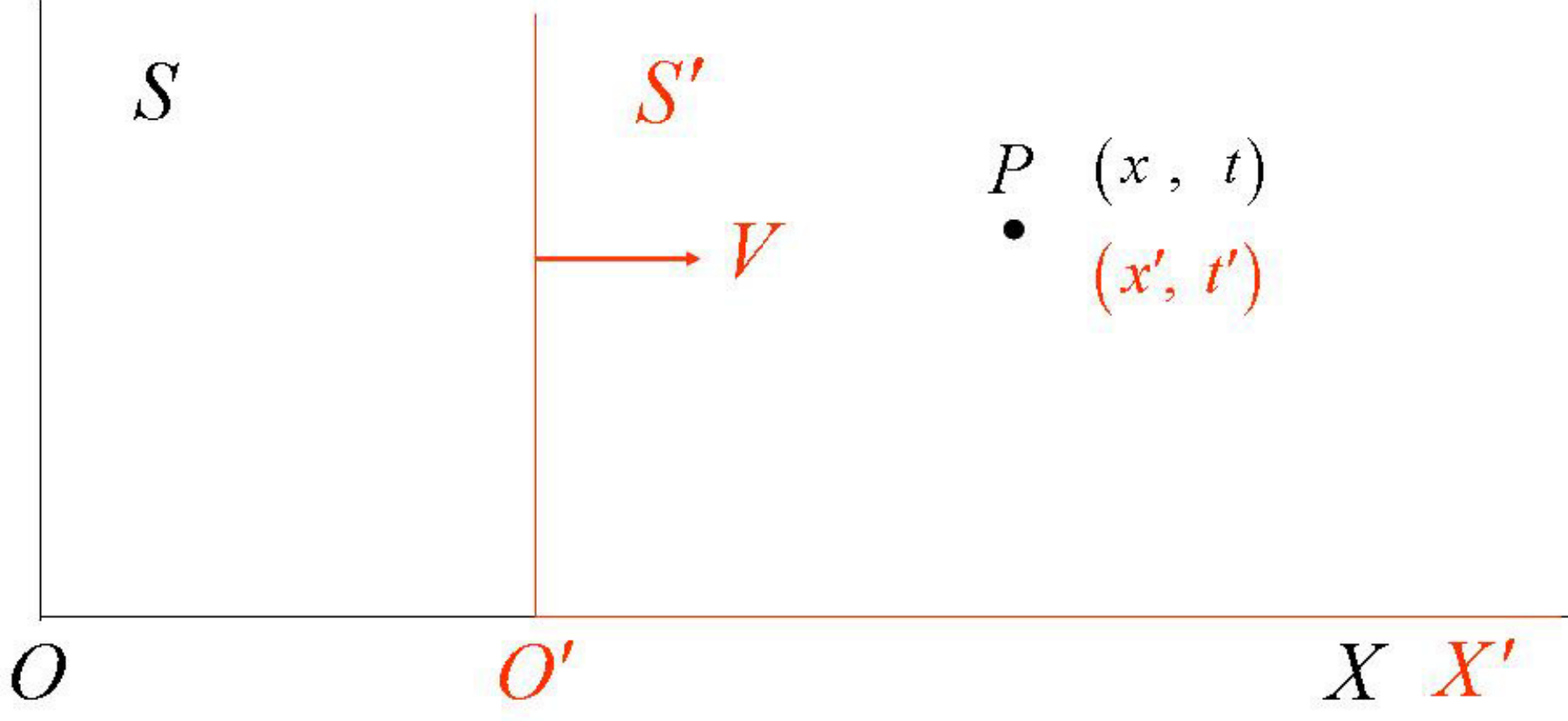}\\
  \caption{ \label{t1}Transformation of space and time for inertial frames}
\end{figure}

Consider an inertial frame $S'$  moves along the $X$ direction with respect to another inertial frame $S$ at relative velocity $V$.  A space-time point $P$ has coordinates $(x',y',z', t')$ in frame $S'$  and coordinates $(x,y,z,t)$ in frame $S$,  as sketched in Figure~\ref{t1}.

The transformation of space and time coordinates should be linear, so that the inverse transform takes the same form.  Most generally, we can write
\begin{equation}\label{1.1}
\left\{\begin{array}{l}
x'=\gamma x+\delta t\\
y'=y\\
z'=z\\
t'=\alpha x+\beta t.
\end{array}\right.
\end{equation}

Consider the motion of the origin $O'$  in frame $S'$:  $x'=0= \gamma x+\delta t$.  It gives $\frac{dx}{dt}=-\frac{\delta}{\gamma}=V$ , so that we have $\delta =-\gamma V$ ; similarly for the origin $O$ in frame $S$:  $x=0$ gives $x'=\delta t, t'=\beta t, \frac{dx'}{dt'}=\frac{\delta}{\beta}=-V $, and we have $\beta =- \frac{\delta}{V}=\gamma$. We have two unknown functions depending on the relative velocity, i.e. $\gamma =\gamma (V), \alpha =\alpha (V)$. Then Eq.~(\ref{1.1}) can be cast into the form

\begin{equation}{\label{1.2}}
\left\{\begin{array}{l}
x'=\gamma (V)( x-V t)\\
y'=y\\
z'=z\\
t'=\alpha (V) x+\gamma (V) t.
\end{array}\right.
\end{equation}
Under spatial reflection,  the first and the fourth equation in Eq.~(\ref{1.2}) change into $x'=\gamma (-V)( x-V t), ~t'=-\alpha (-V) x+\gamma (-V) t$ , so we have
\begin{equation}\label{1.3}
\left\{\begin{array}{l}
\gamma (-V)=\gamma (V)\\
\alpha (-V)=-\alpha (V).
\end{array}\right.
\end{equation}
Alternatively, frame $S$  is moving at velocity $-V$  with respect to frame $S'$ , and space and time transform as the inverse of Eq.~(\ref{1.2})
\begin{equation}{\label{1.4}}
\left\{\begin{array}{l}
x=\gamma (-V)( x'+V t')=\gamma (V)( x'+V t')\\
y=y'\\
z=z'\\
t=\alpha (-V) x'+\gamma (-V) t'=-\alpha (V)+\gamma (V)t'.
\end{array}\right.
\end{equation}
In the trivial case $V=0$ , i.e. the two inertial frames coincide, we must have $x'=x$, $t'=t$. It gives
\begin{equation}\label{1.5}
\left\{\begin{array}{l}
\gamma (V=0)=1\\
\alpha (V=0)=0.
\end{array}\right.
\end{equation}
When Eq.~(\ref{1.2}) and Eq.~(\ref{1.4}) are combined, we obtain $x'=\gamma (V)[\gamma (V)+\alpha(V)V]x'$, which implies
 \begin{equation}\label{1.6}
\alpha (V)V=\frac{1}{\gamma (V)}-\gamma (V).
\end{equation}
Thus we find that functions $\alpha(V)$ and $\gamma (V)$ are related. Once one independent function is specified, the space-time transformation is uniquely determined. Furthermore, by differentiating Eq.~(\ref{1.2}) we find
\begin{equation}{\label{1.7}}
\left\{\begin{array}{l}
dx'=\gamma (d x-V dt)=\gamma (v_x-V)dt\\
dy'=dy=v_ydt\\
dz'=dz=v_zdt\\
dt'=\alpha dx+\gamma dt=\gamma (1+\frac{\alpha}{\gamma}v_x)dt,
\end{array}\right.
\end{equation}
so transformation of velocities between frames is then derived
\begin{equation}{\label{1.8}}
\left\{\begin{array}{l}
v'_x= \frac{dx'}{dt'}=\frac{v_x-V}{1+(\alpha/ \gamma)v_x}
=\frac{v_x-V}{1+(\gamma^{-2}-1)(v_x/V)} \\
\\
v'_y=\frac{dy'}{dt'}=\frac{v_y}{\gamma [1+(\alpha/ \gamma)v_x]}
=\frac{v_y}{\gamma+(\gamma^{-1}-\gamma)(v_x/V)}\\
\\
v'_z=\frac{dz'}{dt'}=\frac{v_z}{\gamma [1+(\alpha/ \gamma)v_x]}
=\frac{v_z}{\gamma+(\gamma^{-1}-\gamma)(v_x/V)}.\\
\end{array}\right.
\end{equation}

This still remains partially unknown, as $\alpha (V)$ or $\gamma (V) $ is yet to be determined.
For the case of  $\alpha (V)=0$ , Eq.~(\ref{1.6}) implies $[\gamma (V)]^2=1 $. To comply with the limit $\gamma (V=0)=1 $, we have $\gamma (V)=1 $. This case is nothing but the Galilean relativity
\begin{equation}{\label{1.9}}
\left\{\begin{array}{l}
x'=x-Vt\\
y'= y \\
z'= z \\
t'= t.
\end{array}\right.
~~~
\left\{\begin{array}{l}
v'_x=v_x-V\\
v'_y=v_y\\
v'_z=v_z.
\end{array}\right.
\end{equation}

\subsection{ \bf Velocity transformation for non-Galilean case}

In this sub-Section, we discuss the non-Galilean case $\alpha (V)\neq 0 $ or $\gamma (V)\neq 1$, which  is more related to the real world when compared to the less interesting trivial case $V=0$. We start with the transformation for velocity and study the kinematic feature it reveals. Assume a particle moving along the $X$ direction with velocity $\vec{v}=(v,0,0)$ in $S$ frame. Now consider a second frame $S'$  moving relative to $S$ frame along the $X$ direction with velocity $V$. Since all motions are along the $X$ direction, from now on for compactness, we drop the subscript $x$ that denotes the Cartesian component of the velocity. The velocity transformation reads
\begin{equation}\label{10}
v'=f(v,V)=\frac{v-V}{1+\frac{\alpha (V)}{\gamma (V)}v}.
\end{equation}
From Eq.~(\ref{1.6}) we furthermore obtain
\begin{equation}\label{11}
1+\frac{\alpha (V)}{\gamma (V)}V=\frac{1}{[\gamma (V)]^2}>0.
\end{equation}
The partial derivative of $f(v,V)$ with respect to the velocity $v$ is positive definite, i.e.
\begin{equation}\label{12}
\frac{\partial f(v,V)}{\partial v}=\frac{1+\frac{\alpha (V)}{\gamma (V)}V}{[1+\frac{\alpha (V)}{\gamma (V)}v]^2}=\frac{1}{[\gamma (V)+\alpha (V)v]^2}>0.
\end{equation}
Namely, the velocity transformation $v'=f(v,V)$ is monotonically increasing with respect to $v$, in either of its continuous ranges: (I) $v\in (-\infty, -\frac{\gamma (V)}{\alpha (V)})$ or (II) $v\in ( -\frac{\gamma (V)}{\alpha (V)}, \infty)$.  For a given value of $V$, we find the convexity from the sign of its second derivative
\begin{equation}\label{13}
\frac{\partial^2 f(v,V)}{\partial v^2}=\frac{-2\alpha (V)} {[\gamma (V)+\alpha (V)v]^3}.
\end{equation}
It can be seen that in range (I), $v'=f(v,V)$ is monotonically increasing and concave, because from Eq.~(\ref{13}) it follows  $v<-\frac{\gamma(V)}{\alpha (V)}$ that $\frac{\partial^2 f(v,V)}{\partial v^2}>0$, for either $\alpha (V)<0$ or $\alpha (V)>0$. On the contrary, in range (II), $v'=f(v,V)$ is monotonically increasing but convex, because the same analysis will conclude $\frac{\partial^2 f(v,V)}{\partial v^2}<0$.

From Eq.~(\ref{11}) we can write: $-\frac{\alpha (V)}{\gamma (V)}V=1-\frac{1}{[\gamma (V)]^2}$, thus $-\frac{\gamma (V)}{\alpha (V)}$ and $V$ have the same sign for $\gamma (v)>1$; while for $\gamma (V)<1$, $-\frac{\gamma (V)}{\alpha (V)}$ and $V$ have opposite signs. Now let us take a closer look at each of the possibilities for case $\gamma (V)\neq 1$:

 (a) $\gamma (V)>1, V>0$, namely $-\frac{\gamma (V)}{\alpha (V)}$ and $V$ have the same positive sign, we then have $0<V<-\frac{\gamma (V)}{\alpha (V)}$.

In range (I), i.e. $v\in (-\infty, ~-\frac{\gamma (V)}{\alpha (V)})$, the behavior of $f(v,V)$ at some special points are given as:
\begin{equation}\label{14}
\left\{\begin{array}{l}
f(v=-\infty,~V)=\frac{\gamma (V)}{\alpha (V)}<0\\
f(v=0, ~V)=-V<0\\
f(v=V, ~V)=0\\
f(v=-\frac{\gamma (V)}{\alpha (V)}, ~V)=\infty.
\end{array}\right.
\end{equation}

\begin{figure}[ht]
  \includegraphics[scale=0.4]{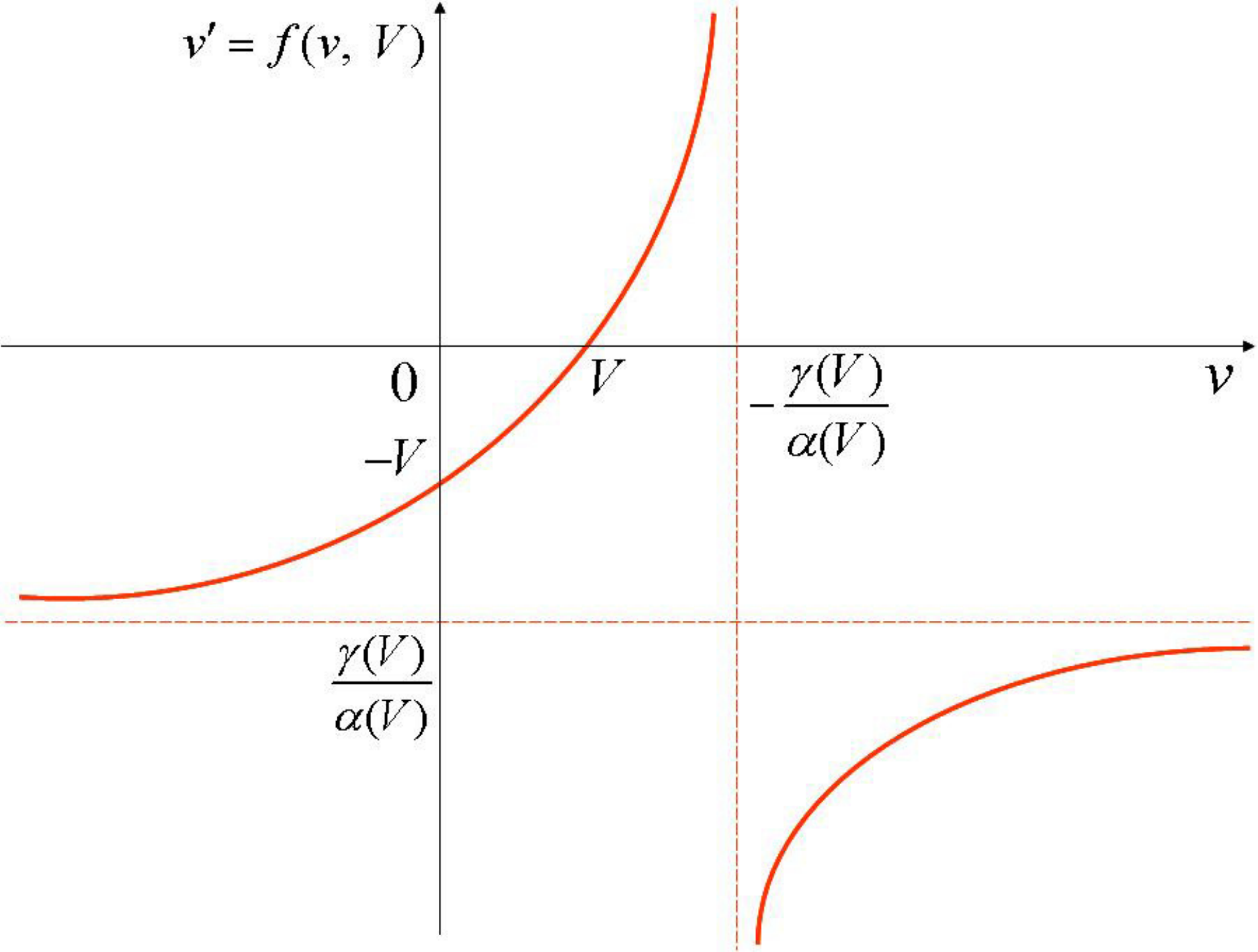}\\
  \caption{\label{fig:t2}Sketches of the velocity transformation: $0<V<-\frac{\gamma (V)}{\alpha (V)}$}
\end{figure}

Similarly, in range (II), i.e. $v\in ( -\frac{\gamma (V)}{\alpha (V)},~\infty)$, we find
\begin{equation}\label{15}
\left\{\begin{array}{l}
f(v=-\frac{\gamma (V)}{\alpha (V)}, ~V)=-\infty\\
f(v=\infty, ~V)=\frac{\gamma (V)}{\alpha (V)}<0.
\end{array}\right.
\end{equation}
The general behavior of $v'=f(v,V)$ is plotted in Figure~\ref{fig:t2}, with asymptotic lines at the point of discontinuity $-\frac{\gamma (V)}{\alpha (V)}$.

(b) $\gamma (V)>1, V<0$, namely $-\frac{\gamma (V)}{\alpha (V)}$ and $V$ have the same negative sign, we then obtain $-\frac{\gamma (V)}{\alpha (V)}< V <0$.

In range (I), we have
\begin{equation}\label{16}
\left\{\begin{array}{l}
f(v=-\infty, ~V)= \frac{\gamma (V)}{\alpha (V)}>0\\
f(v=-\frac{\gamma (V)}{\alpha (V)}, ~V)=\infty.
\end{array}\right.
\end{equation}

\begin{figure}[ht]
  \includegraphics[scale=0.4]{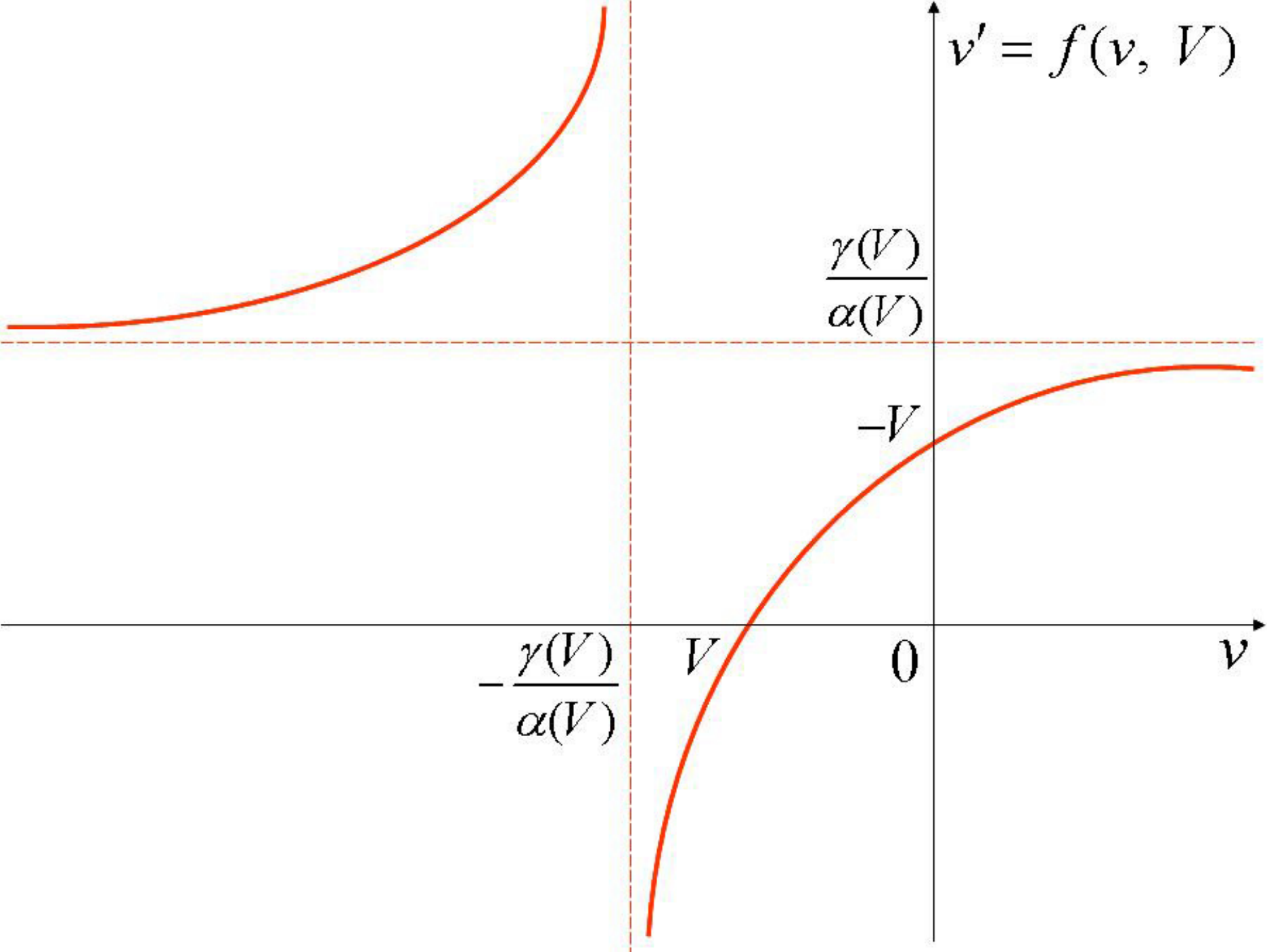}\\
  \caption{ \label{fig:t3}$ -\frac{\gamma (V)}{\alpha (V)}<V<0$}
\end{figure}

And in range (II), we have
\begin{equation}\label{17}
\left\{\begin{array}{l}
f(v=-\frac{\gamma (V)}{\alpha (V)}, ~V)=-\infty\\
f(v=V, ~V)=0\\
f(v=0, ~V)=-V>0\\
f(v=\infty, ~V)=\frac{\gamma (V)}{\alpha (V)}>0.
\end{array}\right.
\end{equation}
In this case $v'=f(v,V)$ is plotted in Figure~\ref{fig:t3}.

(c) $\gamma (V)<1, V>0$, namely $-\frac{\gamma (V)}{\alpha (V)}$ and $V$  have the opposite sign, we then obtain  $-\frac{\gamma (V)}{\alpha (V)}<0 <V$.

In range (I), we have
\begin{equation}\label{18}
\left\{\begin{array}{l}
f(v=-\infty, ~V)= \frac{\gamma (V)}{\alpha (V)}>0\\
f(v=-\frac{\gamma (V)}{\alpha (V)}, ~V)=\infty.
\end{array}\right.
\end{equation}

\begin{figure}[ht]
  \includegraphics[scale=0.4]{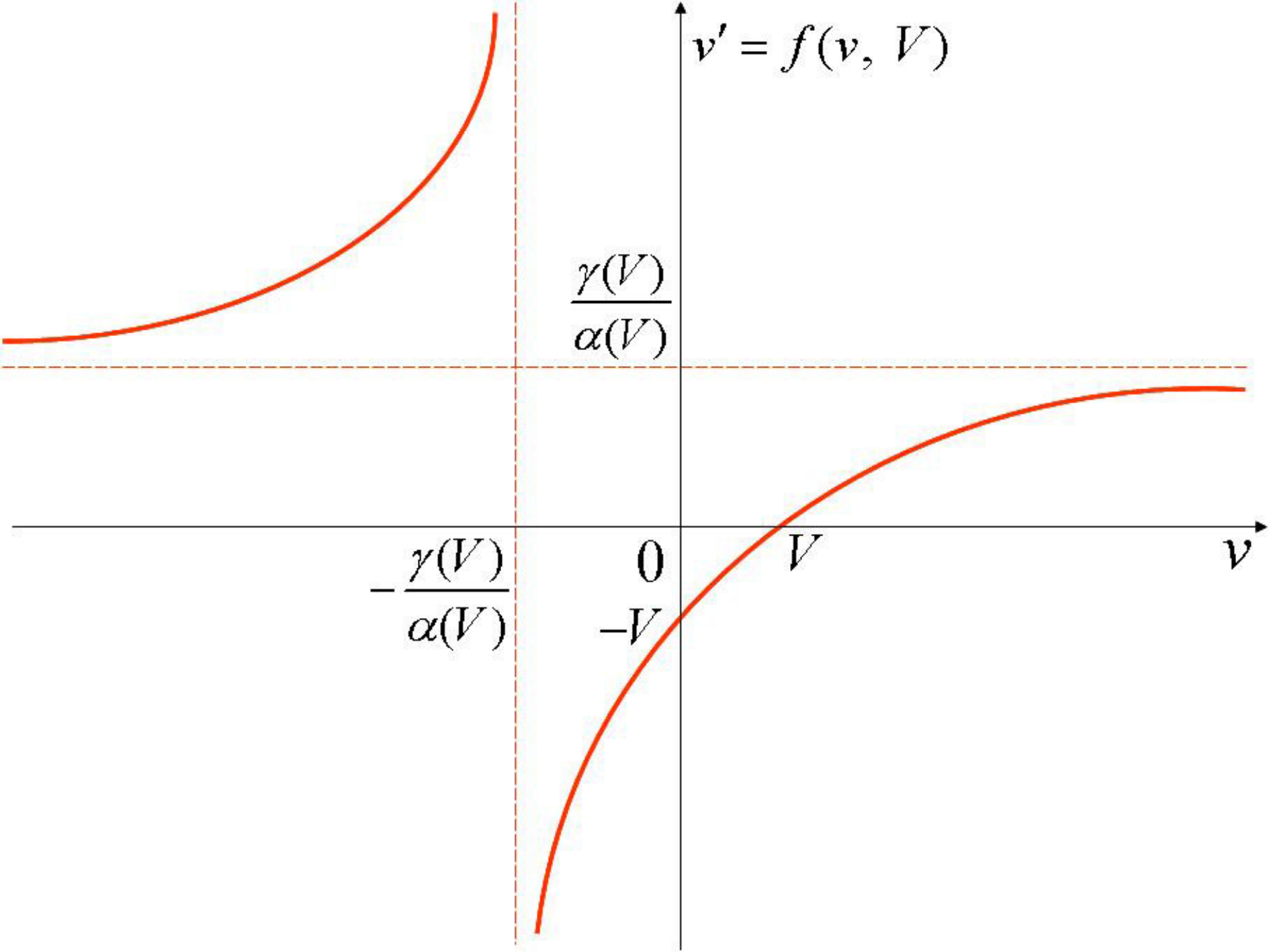}\\
  \caption{ \label{fig:t4}$ -\frac{\gamma (V)}{\alpha (V)}<0<V$}
\end{figure}

And in range (II), we have
\begin{equation}\label{19}
\left\{\begin{array}{l}
f(v=-\frac{\gamma (V)}{\alpha (V)}, ~V)=-\infty\\
f(v=0, ~V)=-V<0\\
f(v=V, ~V)=0\\
f(v=\infty, ~V)=\frac{\gamma (V)}{\alpha (V)}>0.
\end{array}\right.
\end{equation}
Correspondingly $v'=f(v,V)$ is plotted in Figure~\ref{fig:t4}.

(d) $\gamma (V)<1, V<0$, namely $-\frac{\gamma (V)}{\alpha (V)}$ and $V$ have the opposite sign, we then obtain $V<0< -\frac{\gamma (V)}{\alpha (V)}$.

In range (I), we have
\begin{equation}\label{20}
\left\{\begin{array}{l}
f(v=-\infty, ~V)= \frac{\gamma (V)}{\alpha (V)}<0\\
f(v=V, ~V)=0\\
f(v=0, ~V)=-V>0\\
f(v=-\frac{\gamma (V)}{\alpha (V)}, ~V)=\infty.
\end{array}\right.
\end{equation}

\begin{figure}[ht]
  \includegraphics[scale=0.4]{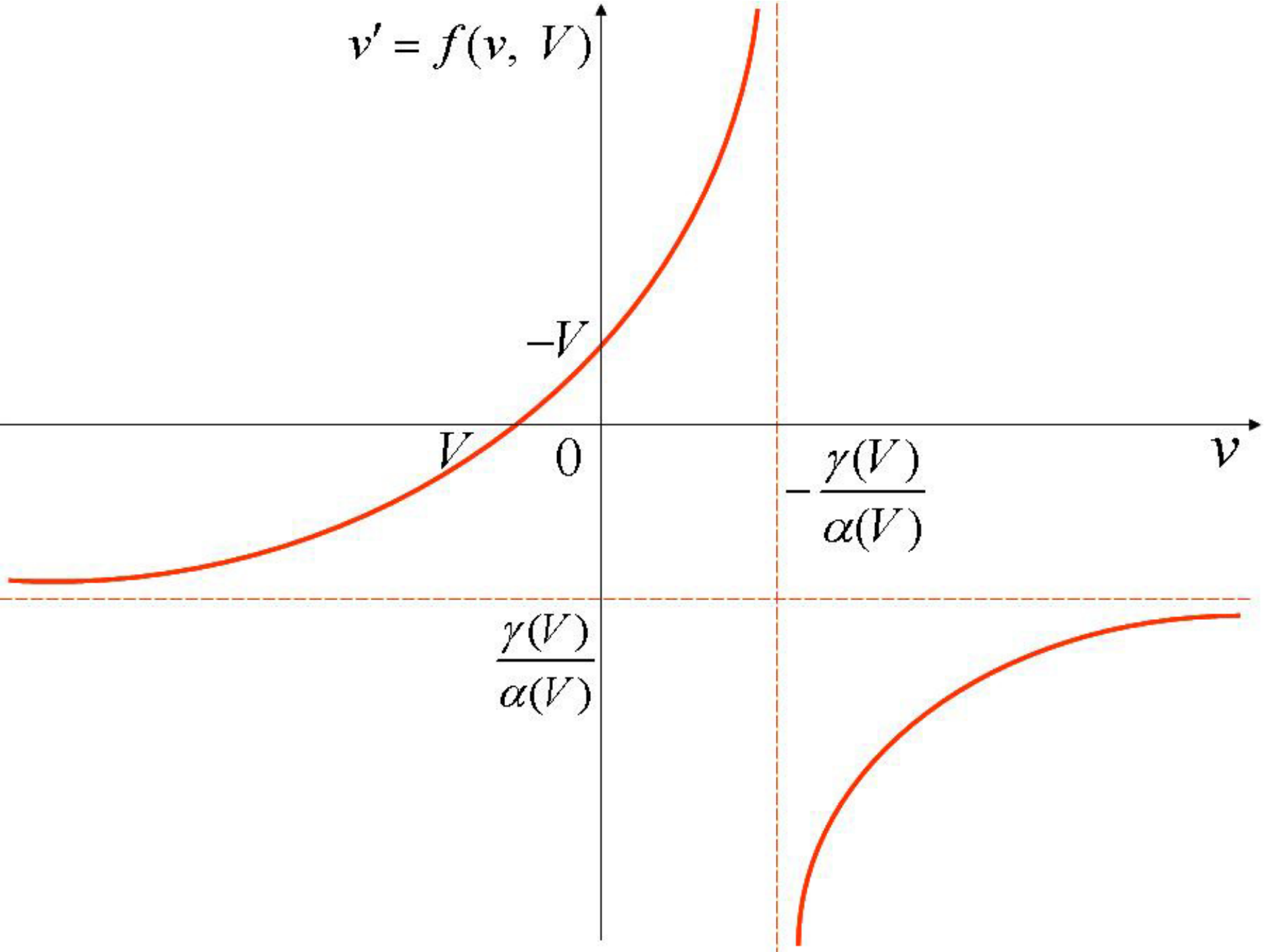}\\
  \caption{\label{fig:t5}$ V<0 <-\frac{\gamma (V)}{\alpha (V)}$ }
\end{figure}

And in the range (II), we have
\begin{equation}\label{21}
\left\{\begin{array}{l}
f(v=-\frac{\gamma (V)}{\alpha (V)}, ~V)=-\infty\\
f(v=+\infty, ~V)=\frac{\gamma (V)}{\alpha (V)}<0.
\end{array}\right.
\end{equation}
In this case $v'=f(v,V)$ is plotted in Figure~\ref{fig:t5}.

To determine the allowed range for the particle's velocity $v$, we consider two categories depending on whether a rest frame for the particle exists or not.

(i)Particles having no rest frames

For arbitrary relative velocity $V$, we would have $v'=f(v,V)\neq 0$, i.e. $v-V\neq 0$. In this case, $V$ is not  allowed to equal to the particle's velocity, while in principle the velocity can pass the discontinuity point $v=-\frac{\gamma (V)}{\alpha (V)}$. At the discontinuity point we have $|v'|=\infty$. Since the particle has no rest frame, we have $0<|v'|$, $0<|v|$, so that the allowed value for $|v|$ is bounded from below with $v_{min}$, namely $0<v_{min}<|v|<\infty$. Therefore, for particles having no rest frames, there is a lower bound for its speed.

(ii)Particles having rest frames

For arbitrarily allowed velocity $v\neq 0$, a rest frame for the particle can always be found, to which a transformation of $V\neq 0$ finds the particle stationary $v'=f(v,V)= 0$, i.e.
\begin{equation}\label{22}
\left\{\begin{array}{l}
v-V=0\\
1+\frac{\alpha (V)}{\gamma (V)}v\neq 0.
\end{array}\right.
\end{equation}
The above equations require $v=V, v\neq -\frac{\gamma (V)}{\alpha (V)}$, which implies that for arbitrary $V\neq 0$, the velocity transformation $v'=f(v,V)$ is free from singularity at the discontinuity point  $-\frac{\gamma (V)}{\alpha (V)}$. Accordingly, the velocity transformation  $v'=f(v,V)$ is continuous throughout and monotonically increasing (the continuous curve passing $v=V$ , as shown in Figure~\ref{fig:t2}, \ref{fig:t3}, \ref{fig:t4} and \ref{fig:t5}.

\begin{figure}[ht]
  \includegraphics[scale=0.4]{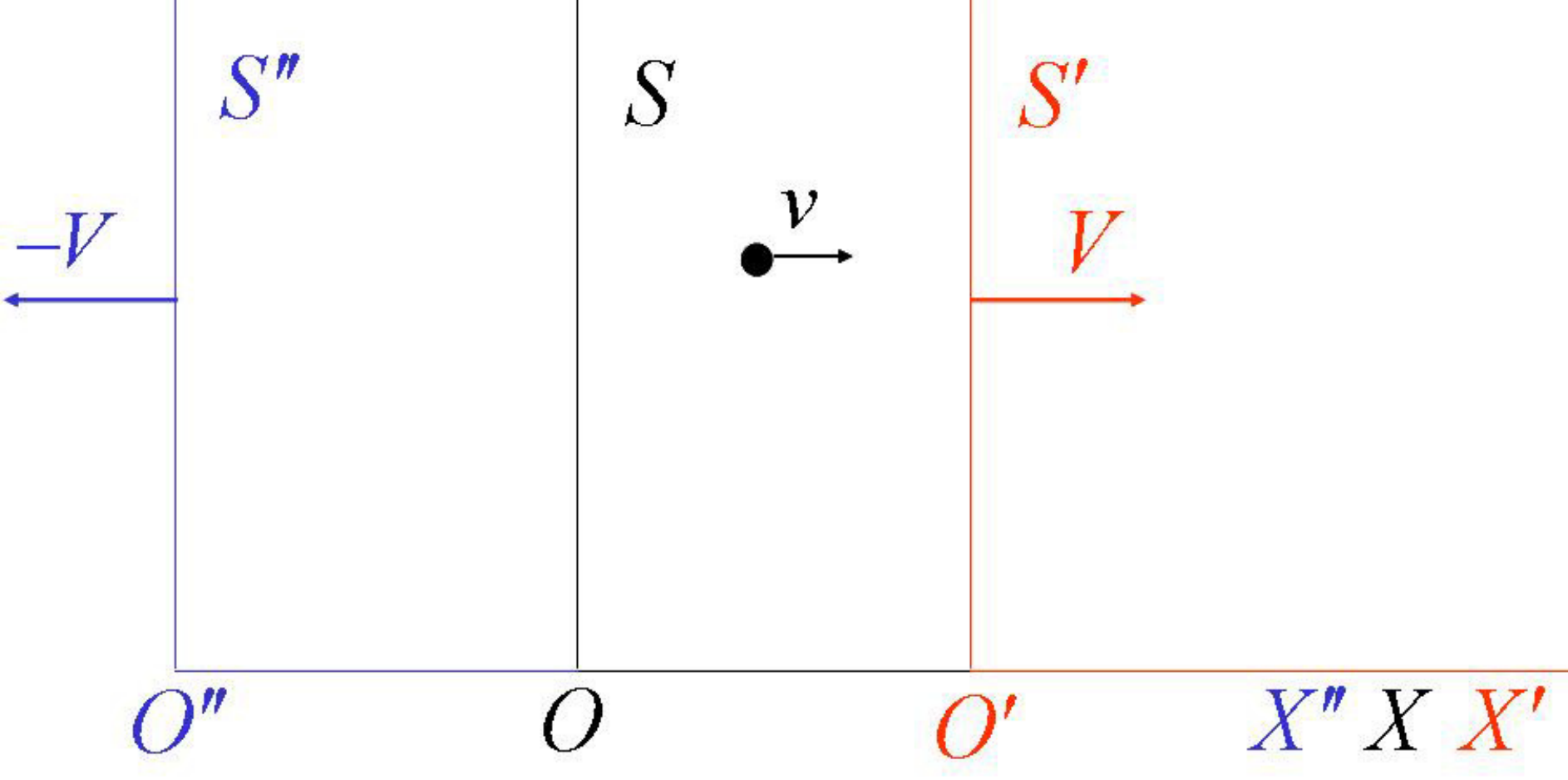}\\
  \caption{Transformation of velocity under boosts} \label{fig:t6}
\end{figure}

Now consider a particle moving along the $X$ direction in inertial frame $S$, with velocity $\vec{v}=(v,0,0)$. Consider two new frames, frame $S'$ moves along the $X$ direction at velocity $V>0$ with respect to frame $S$, and frame $S''$ along the $X$ direction at velocity $-V$ with respect to frame $S$ (i.e. frame $S$ moves at $V>0$ relative to frame $S''$ ) as shown in Figure~\ref{fig:t6}.
Since the transformation from $S''$ to $S$ and the one from $S$ to $S'$ are identical, from Eq.~(\ref{10}) and Eq.~(\ref{1.3}), we can write down the relations between the particle's velocities as measured in all three inertial frames
\begin{equation}\label{23}
\left\{\begin{array}{l}
v'=f(v,~V)=\frac{v-V}{1+[\alpha (V)/\gamma (V)]v}\\
v=f(v',-V)=\frac{v'+V}{1-[\alpha (V)/\gamma (V)]v'}\\
v=f(v'',~V)=\frac{v''-V}{1+[\alpha (V)/\gamma (V)]v''}\\
v''=f(v,-V)=\frac{v+V}{1-[\alpha (V)/\gamma (V)]v}.
\end{array}\right.
\end{equation}

Recall that $f(v,V)$  monotonically increases with respect to $v$, so if $v\leq v''$, then we have $f(v,V)\leq f(v'',V)$ , which converts to $v'\leq v$ ; and {\it vice versa}. Therefore, we find either $v'\leq v\leq v''$ or $v'\geq v\geq v''$ holds, which is a general self-consistency condition that any velocity transformation rules (including the Galilean and the Lorentz transformation) should satisfy.

We now falsify by absurdity that for non-Galilean transformation particles having rest frames cannot travel at arbitrarily large speed: assume no speed limit exists for a certain particle species, i.e. it is always possible to accelerate the particle to arbitrary extent $|v|\rightarrow\infty$ in a certain frame $S$. Because of $\alpha (V)\neq 0$, finite velocity will be observed in either frame $S'$  or frame $S''$
\begin{equation}\label{24}
\left\{\begin{array}{l}
v'=\lim_{v\rightarrow\pm\infty}~ \frac{v-V}{1+[\alpha (V)/\gamma (V)]v}= \frac{\gamma (V)}{\alpha (V)}\\
\\
v''=\lim_{v\rightarrow\pm\infty}~ \frac{v+V}{1-[\alpha (V)/\gamma (V)]v}=-\frac{\gamma (V)}{\alpha (V)}=-v'.
\end{array}\right.
\end{equation}

It can then be derived that $|v'|=|v''|=|\frac{\gamma (V)}{\alpha (V)}|<|v|$. However, this contradicts with the fact that $|v|\leq |v'|=|v''|$  at $v'=-v''$, regardless of $v'\leq v\leq v''$ or $v'\geq v\geq v''$. Remarkably, we have falsified the previous assumption $|v|\rightarrow\infty$  that we have taken for granted. In other words, we have proved that given a consistent space-time transformation for any particle having rest frames, their cannot be accelerated to infinite speed, but are bounded from above by a speed limit $v_{max}$, i.e. $0\leq |v|<v_{max}$. The only exception is the Galilean transformation with $\alpha (V)=0$ and $\gamma (V)=1$, which admits an absolute division in notion between space and time.

For particles admitting rest frames, a frame transformation always exists for all allowed values of $v$  so that  $v'=f(v,V)=0$, i.e. $v=V$.  Given the constraint $ 1+\frac{\alpha (V)}{\gamma (V)}V>0 $ on $V$ from Eq.(\ref{11}), the velocity should also be subject to the constraint
\begin{equation}\label{25}
1+\frac{\alpha (v)}{\gamma (v)}v>0.
\end{equation}
followed by that a rest frame with $v=V$  exists. Consider the case in which $-\frac{\alpha (v)}{\gamma (v)}$ and  $v$  have the same sign, because of $-1<\frac{\alpha (v)}{\gamma (v)}v <0<1$, i.e. $ |\frac{\alpha (v)}{\gamma (v)}v |<1 $, we infer that $|v|$ is bounded by $|v|<|\frac{\gamma (v)}{\alpha (v)}|$. A familiar example would be the Lorentz transformation, in which $\gamma (V)=(1-\frac{V^2}{c^2})^{-1/2}$ and $\frac{\alpha (V)}{\gamma (V)}=-\frac{V }{c^2} $. From $1+\frac{\alpha (V)}{\gamma (V)}V>0$ we find constraint $|V|<c$, and using $1+\frac{\alpha (v)}{\gamma (v)}v>0$  we furthermore have $|v|<|\frac{\gamma(v)}{\alpha (v)}|=\frac{c^2}{|v|}$ £¬i.e. $|v|<c$.

We have shown that the particles having rest frames are subject to a upper speed limit $v_{max}$  such that $|\vec{v}|<v_{max}$ , with the only exception being the Galilean relativity. Now consider the possibility that certain particles actually exist in Nature that saturate the upper limit speed  $v=v_{max}$. Recall Figure~\ref{fig:t6} and the inequality $v'\leq v\leq v''$  or $v'\geq v\geq v''$  , we can infer: (i) on one hand we have $v''\geq v= v_{max}$ , and on the other hand by definition of the upper limiting speed we have $v''\leq v_{max}$ , so we must have $v''= v_{max}$. Furthermore from Eq.~(\ref{23}), $v''=\frac{v_{max}+V}{1-[\alpha (V)/\gamma (V)]v_{max}}=v_{max}$, it can be solved that $\frac{\alpha (V)}{\gamma (V)}=-\frac{V}{v^2_{max}}$. (ii) if on the contrary  $v'\geq v= v_{max}$  , the same argument forces to have $v'=v_{max}$ , and then from Eq.~(\ref{23}) we obtain $v'=\frac{v_{max}-V}{1+[\alpha (V)/\gamma (V)]v_{max}}=v_{max}$, so again we obtain $\frac{\alpha (V)}{\gamma (V)}=-\frac{V}{v^2_{max}}$. Then follows from $\alpha (V) V=\frac{1}{\gamma (V)}-\gamma (V)$, the non-Galilean transformation is obtained
\begin{equation}\label{26}
\left\{\begin{array}{l}
 \gamma (V)=\frac{1}{\sqrt{1-\frac{V^2}{v_{max}^2}}}\\
 \alpha (V)=-\frac{V}{v_{max}^2}\frac{1}{\sqrt{1-\frac{V^2}{v_{max}^2}}}.
\end{array}\right.
\end{equation}

We now investigate invariant quantities in the velocity transformation. Assume $v=v_0$ is a fixed point in the velocity transformation Eq.~(\ref{10}), for arbitrary $V\neq 0$, namely
\begin{equation}\label{27}
v'= \frac{v_0-V}{1+\frac{\alpha (V)}{\gamma (V)}v_0}=v_0,
\end{equation}
then we get
\begin{equation}\label{28}
v_0^2=-\frac{\gamma (V)}{\alpha (V)}V>0.
\end{equation}
This equation states that an invariant speed $v_0$  exists only if $ -\frac{\gamma (V)}{\alpha (V)}$  and $V$ have the same sign i.e. $\gamma (V) >1$ , satisfied $0<V<-\frac{\gamma (V)}{\alpha (V)}$ in Figure~\ref{fig:t2} or $ -\frac{\gamma (V)}{\alpha (V)}<V<0$  in Figure~\ref{fig:t3}. Such an invariant velocity does not exist in the case $\gamma (V)<1$ , where $ -\frac{\gamma (V)}{\alpha (V)}$  and $V$ have the opposite signs as in Figure~\ref{fig:t4} and Figure~\ref{fig:t5}.  Also, the Galilean transformation with $\alpha (V)=0$  does not admit invariant velocity. Imagine that there are particle species in Nature that move at invariant speed $v_0$  (e.g. the assumption of constant speed of light in the conventional formulation of Special Relativity), then from Eq.~(\ref{27}) we obtain
\begin{equation}\label{29}
 \frac{\alpha (V)}{\gamma (V)}=-\frac{V}{v_0^2}.
\end{equation}
It then follows from Eq.~(\ref{1.6}) and the condition $\gamma(V=0)=1$ that
\begin{equation}\label{30}
\left\{\begin{array}{l}
 \gamma (V)=\frac{1}{\sqrt{1-\frac{V^2}{v_0^2}}}\\
 \alpha (V)=-\frac{V}{v_0^2}\frac{1}{\sqrt{1-\frac{V^2}{v_0^2}}}.
\end{array}\right.
\end{equation}

Compare Eq.~(\ref{30}) with Eq.~(\ref{26}), it is concluded that the upper speed limit is equal to invariant speed $v_{max}=v_0$. It is worth pointing out that we have only been able to show that particle velocity can be arbitrarily close to the limiting value $|\vec{v}|\rightarrow v_{max}$  in principle, but it should not be taken for granted that the bound $v_{max}$  can actually be saturated. Clearly it is unjustified to set $|\vec{v}|=v_{max}=v_0$ in the velocity transformation to derive $\gamma (V)$ and $\alpha (V)$. Namely, without new theoretical assumptions, the exact form of $\gamma (V)$ and $\alpha (V)$ cannot be determined only from the knowledge of kinematics~\cite{9,10}.


\section{Special relativistic dynamics}
\label{sec:dynamics}

The traditional relativistic dynamics is formulated using the Lorentz transformation of the relativistic kinematics~\cite{13}. In this Section, we present a new derivation of the mass-velocity and mass-energy relations solely within the framework of relativistic dynamics, which does not necessitate any explicit form of space-time transformation. then the generalized Lorentz transformations for space-time is derived. In our new formula, the Relativity Constant $C_R$, required by the principle of relativity, replaces the speed of light $c$, so mass-velocity and mass-energy relations as well as the generalized Lorentz transformations are not determined by the physical properties of the photon. The upper speed limit for non-zero rest mass tardyons with $|\vec{v}|<C_R$, and the lower speed limit for real mass tachyons with $|\vec{v}|>C_R$, are both obtained naturally by solving the mass-velocity differential equation, with no need of the assumption of constant speed of light.

\subsection{\bf Particles having rest frames}

In this sub-Section, our discussion are restricted to particle species which have rest frames.

For a particle with intrinsic mass $m_0> 0$, we consider its momentum $\vec{p}=\vec{p}(m_0, \vec{v})$ at a velocity of $\vec{v}$. It is reasonable to have $\vec{p}(m_0, \vec{v}=0)=0$ when the particle is at rest $\vec{v}=0$. Under spatial reflection $\vec{r}\rightarrow -\vec{r}$ or time reflection $t \rightarrow -t$, we have $\vec{v}=\frac{d\vec{r}}{dt}\rightarrow -\vec{r}$ and $\vec{p}\rightarrow -\vec{p}$. Particle's momentum must be proportional to the intrinsic mass as it has to be additive. From dimensional analysis, we parameterize the momentum as $\vec{p}=m_0g(\vec{v})\vec{v}$, where the dimensionless function $g(\vec{v})$ satisfies $g(-\vec{v})=g(\vec{v})$ and stays invariant under three-dimensional spatial rotations. It can only be that $g(\vec{v})=g(v^2)$, and we introduce the notation
\begin{equation}\label{31}
m=m(v^2)=m_0g(v^2).
\end{equation}
The relativistic momentum then can be most generally defined as
\begin{equation}\label{32}
\vec{p}=m_0g(v^2)\vec{v}=m\vec{v}.
\end{equation}
For those particle species which have rest frames, the intrinsic mass $m_0$ are usually called rest mass and $m$ called moving mass respectively~\cite{14}. The mass-velocity relation $m=m(v^2)$ is yet to be determined. In the limit $\vec{v}\rightarrow 0$, we would have $m_0=m(0)=m_0g(0)$, followed by the condition $g(0)=1$. On the other hand, we immediately recover the Newtonian momentum in the small-velocity limit $\vec{p}\rightarrow m_0\vec{v}$. Given the relativistic definition of momentum $\vec{p}=m\vec{v}$, the external force $\vec{F}$ exerted on the particle can be then defined through
\begin{equation}\label{33}
d\vec{p}=\vec{F}dt.
\end{equation}
And the particle's relativistic energy $E$ can be then defined through
\begin{equation}\label{34}
dE=\vec{F}\cdot d\vec{r}.
\end{equation}
It is thus made apparent that the total relativistic momentum and energy are conserved separately in an isolated system~\cite{15}.

\begin{figure}[ht]
  \includegraphics[scale=0.4]{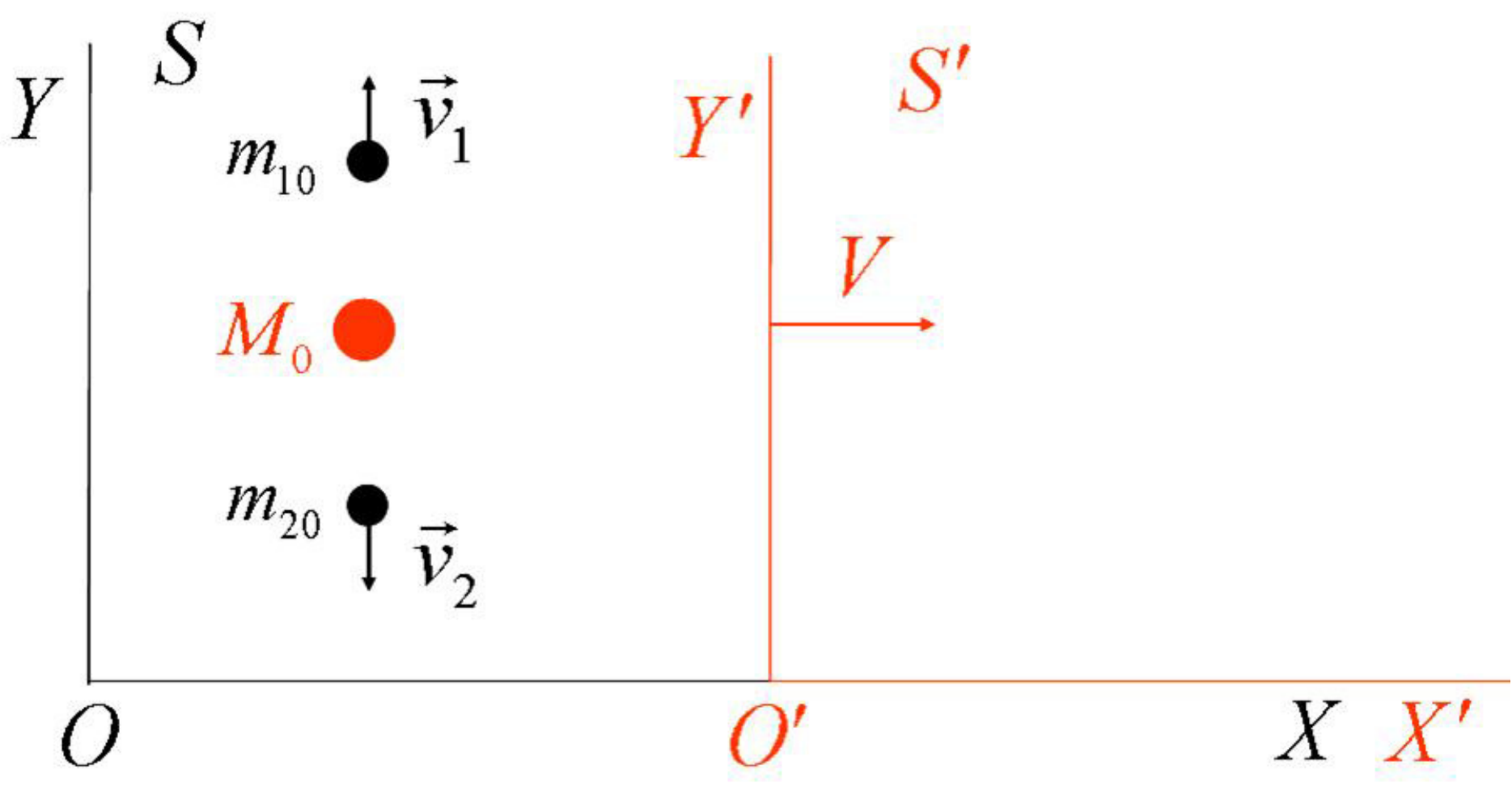}\\
  \caption{ Particle decay as observed in different inertial frames} \label{fig:t7}
\end{figure}

By similar argument based on dimensional analysis and the invariance of $E$ under three-dimensional spatial rotations,  the energy can also be most generally parameterized in the following way
\begin{equation}\label{35}
E=E(v^2)=m_0G(v^2).
\end{equation}
Here $G(v^2)$ is another function to be determined. To derive unknown functions $g(v^2)$ and $G(v^2)$ , we proceed by considering a spontaneous decay of a particle with rest mass $M_0$  into two particles with rest masses $m_{10}$ and $m_{20}$ respectively. Take frame $S$ to be the rest frame of the particle $M_0$, with null velocity $\vec{U}=(0,~0,~0)$. After the decay, the two decay products move along the $Y$ direction with velocity $\vec{v_1}=(0,~v_1,~0)$ and $\vec{v_2}=(0, -v_2,~0)$ respectively, where $v_1=|\vec{v_1}|$ and $v_2=|\vec{v_2}|$, as shown in Figure~\ref{fig:t7}. It can be deduced from momentum conservation in the $Y$ direction that
\begin{equation}\label{36}
m_1(v_1^2)v_1- m_2(v_2^2)v_2=0.
\end{equation}
In the special case $m_{10}=m_{20}$, we have $g(v_1^2)v_1=g(v_2^2)v_2$, i.e. $v_1=v_2$, which is obviously true since the two decay products move away from each other at equal speed.

The energy of the particle $M_0$ before it decays is $E=M_0G(0)$, while after decay the products $m_{10}$ and $m_{20}$ have energies $E_1=m_{10}G(v_1^2)$ and $E_2=m_{20}G(v_2^2)$, respectively. Then from conservation of energy we can write
\begin{equation}\label{37}
M_0G(0)=m_{10}G(v_1^2)+m_{20}G(v_2^2).
\end{equation}

Next introduce frame $S'$ which moves along the $X$ direction at speed $V$ with respect to frame $S$, as illustrated in Figure~\ref{fig:t7}. From the general velocity transforming rules in Eq.~(\ref{1.8}), in frame $S'$, the particle $M_0$ has velocity $\vec {V'}=(-V,~0,~0)$ before it decays. After the decay, the particle $m_{10}$ has velocity $\vec {v}'_1=(-V,~ \frac{v_1}{\gamma(V)},~0)$ and the particle $m_{20}$ has velocity $\vec {v}'_2=(-V,~\frac{-v_2}{\gamma(V)}, ~0)$. We therefore obtain
\begin{equation}\label{38}
\left\{\begin{array}{l}
 V'^2=V^2\\
 {v'_1}^2=V^2+\frac{v_1^2}{[\gamma(V)]^2}\\
 {v'_2}^2=V^2+\frac{v_2^2}{[\gamma(V)]^2}.
\end{array}\right.
\end{equation}

According to the principle of relativity, conservation of momentum holds equally well in frame $S'$. Along the $X$ direction, it allows us to write
\begin{equation}\label{39}
M(V'^2)V'_x=m_1({v'_1}^2)v'_{1x}+m_2({v'_2}^2)v'_{2x}.
\end{equation}
Because $V'_x=v'_{1x}=v'_{2x}=-V$, we obtain
\begin{equation}\label{40}
M(V'^2)=m_1({v'_1}^2)+m_2({v'_2}^2).
\end{equation}
On the other hand, in frame $S'$, the particle $M_0$ has energy $E(V'^2)$ before it decays, while after the decay the products $m_{10}$ and $m_{20}$ have energies $E_1({v'_1}^2)$ and $E_2({v'_2}^2)$, respectively. From energy conservation in frame $S'$, we have
\begin{equation}\label{41}
E(V'^2)=E_1({v'_1}^2)+E_2({v'_2}^2).
\end{equation}
Combining that with Eq.~(\ref{40}) and Eq.~(\ref{41}), we furthermore obtain
\begin{equation}\label{42}
\frac {E(V'^2)}{M(V'^2)}=\frac {E_1({v'_1}^2)+E_2({v'_2}^2)}{m_1({v'_1}^2)+m_2({v'_2}^2)}.
\end{equation}
or explicitly,
\begin{equation}\label{43}
\frac {M_0G(V'^2)}{M_0g(V'^2)}=\frac {m_{10}G({v'_1}^2)+m_{20}G({v'_2}^2)}{m_{10}g({v'_1}^2)+m_{20}g({v'_2}^2)}.
\end{equation}
Now let
\begin{equation}\label{44}
A(v^2)=\frac {G(v^2)}{g(v^2)}.
\end{equation}
Then we find
\begin{equation}\label{45}
A(V'^2)=\frac {m_{10}A({v'_1}^2)+m_{20}\frac{g({v'_2}^2)}{g({v'_1}^2)}A({v'_2}^2)}{m_{10}+m_{20}\frac{g({v'_2}^2)}{g({v'_1}^2)} }.
\end{equation}
Taking the special case $m_{10}=m_{20}$, we have $v_1=v_2$ from Eq.~(\ref{36}) and ${v'_1}^2={v'_2}^2$ from Eq.~(\ref{38}). Given these results, Eq.~(\ref{45}) converts into $A(V'^2)=A({v'_1}^2)$. Since $V'^2=V^2$ and ${v'_1}^2=V^2+\frac{v_1^2}{[\gamma(V)]^2}$ should vary independently ( because $V$ and $v_1$ are independent quantities), $A(v^2)$ must be a constant independent of the particle's velocity. Even for the general case $m_{10}\neq m_{20}$, $A(v^2)$ being a constant $A=A(V'^2)=A({v'_1}^2)=A({v'_2}^2)$ is fully consistent with Eq.~(\ref{45}). Therefore, we are justified to write
\begin{equation}\label{46}
G(v^2)=g(v^2)A,
\end{equation}
where the constant $A$ is independent of the particle's velocity. The particle's energy $E(v^2)$ and rest energy $E_0=E(0)$  then read
\begin{equation}\label{47}
\left\{\begin{array}{l}
 E(v^2)=m_0g(v^2)A=m(v^2)A\\
 E_0=m_0A.
\end{array}\right.
\end{equation}
i.e. energy is proportional to the moving mass, which is a necessary condition to enforce the principle of relativity in terms of the forms for energy conservation or mass conservation in all inertial frames. From Eq.~(\ref{33}) and Eq.~(\ref{34}) we have
\begin{equation}\label{48}
dE=\vec{F}\cdot d\vec{r}=\frac{d\vec{p}}{dt}\cdot d\vec{r}=\vec{v}\cdot (md\vec{v}+\vec{v}dm)=\frac{1}{2}md(v^2)+v^2dm=Adm.
\end{equation}
from which we derive a differential equation that the mass-velocity relation solves
\begin{equation}\label{49}
\frac{dm}{m}=\frac{d(v^2)}{2(A-v^2)}.
\end{equation}
Integrating the equation, we find: $\ln m(v^2)=-\frac{1}{2}\ln(A-v^2)+C$ with $C$ being a constant of integration. The logarithmic term requires that $A-v^2 >0$, i.e. $m(v^2)=e^{C}(A-v^2)^{-\frac{1}{2}}$. From boundary condition $m(0)=\frac{e^{C}}{\sqrt{A}}=m_0$, we obtain $e^{C}=m_0\sqrt{A}$ and have
\begin{equation}\label{50}
m=m_0g(v^2)=\frac{m_0}{\sqrt{1-\frac{v^2}{A}}}~~~~(m_0>0~,~v^2<A).
\end{equation}
We have seen from the solving the differential equation for the mass-velocity relation that the condition $v^2<A $ naturally arises without any special assumption and the particle's velocity must has an upper bound. We introduce a constant $C_R$ with the dimension of velocity, which we call the {\it Relativity Constant} (the letter $C$ stands for $Constant$, and the subscript $R$ stands for $Relativity$), so that a particle saturates that bound by achieving $v=|\vec{v}|<C_R $, $(C_R-v)\rightarrow 0 $. Under the action of a constant external force $F$, the particle eventually approaches that speed limit $v \rightarrow C_R $. Throughout the acceleration process, we have
\begin{equation}\label{51}
\int_{v=0}^{C_R }d(mv)=\int_{t=0}^{\infty} Fdt.
\end{equation}
The equation requires that it is feasible to exert a constant external force on the particle in order to accelerate it to $C_R$, which becomes the unique speed limit. However, caveat is put here that particle's velocity does not necessarily reach that speed limit. On the other hand, as in case $v\rightarrow C_R$, we have $\lim m(v)v=F\int_{t=0}^{\infty}dt\rightarrow \infty$, i.e. $m(v\rightarrow C_R)=m_0(1-\frac{C_R^2}{A})^{-\frac{1}{2}}\rightarrow \infty$, from which we fix $A=C_R^2$. Then the mass-velocity relation is eventually derived to be
\begin{equation}\label{52}
m=\frac{m_0}{\sqrt{1-\frac{v^2}{C_R^2}}}~~~~(m_0>0~,~|\vec{v}|< C_R).
\end{equation}
And the mass-energy relation is
\begin{equation}\label{53}
\left\{\begin{array}{l}
 E=\frac{m_0C_R^2}{\sqrt{1-\frac{v^2}{C_R^2}}}=mC_R^2\\
 E_0=m_0C_R^2.
\end{array}\right.
~~~~(m_0>0~,~|\vec{v}|< C_R)
\end{equation}

\subsection{\bf Generalized Lorentz transformation}

Having derived the mass-velocity and mass-energy relations, from Eq.~(\ref{37}): $M_0G(0)=m_{10}G(v_1^2)+m_{20}G(v_2^2)$ and Eq.~(\ref{41}): $M(V'^2)=m_1({v'_1}^2)+m_2({v'_2}^2)$, as well as the relation of Eq.~(\ref{38}), we obtain a set of equations
\begin{equation}\label{54}
\left\{\begin{array}{l}
 M_0=\frac {m_{10}}{\sqrt{1-\frac{v_1^2}{C_R^2}}}+ \frac{m_{20}}{\sqrt{1-\frac{v_2^2}{C_R^2}}}\\
 \frac{1}{\sqrt{1-\frac{V'^2}{C_R^2}}}M_0=\frac{m_{10}}{\sqrt{1-\frac{{v'_1}^2}{C_R^2}}}+ \frac {m_{20}}{\sqrt{1-\frac{{v'_2}^2}{C_R^2}}}.
\end{array}\right.~~~~~~
\left\{\begin{array}{l}
 V'^2=V^2\\
 {v'_1}^2=V^2+\frac{v_1^2}{[\gamma (V)]^2}\\
 {v'_2}^2=V^2+\frac{v_2^2}{[\gamma (V)]^2}.
\end{array}\right.
\end{equation}
which can be further simplified into
\begin{equation}\label{55}
 \frac {m_{10}}{\sqrt{1-\frac{V^2}{C_R^2}-\frac{v_1^2}{C_R^2}(1-\frac{V^2}{C_R^2})}}+ \frac {m_{20}}{\sqrt{1-\frac{V^2}{C_R^2}-\frac{v_2^2}{C_R^2}(1-\frac{V^2}{C_R^2})}}
 = \frac {m_{10}}{\sqrt{1-\frac{V^2}{C_R^2}-\frac{v_1^2}{C_R^2}\frac{1}{[\gamma (V)]^2}}}+ \frac {m_{20}}{\sqrt{1-\frac{V^2}{C_R^2}-\frac{v_2^2}{C_R^2}\frac{1}{[\gamma (V)]^2}}}.
\end{equation}
It can be solved that $[\gamma (V)]^{-2}=1-\frac{V^2}{C_R^2}$, and applying relation Eq.(6): $\alpha (V)V=\frac{1}{\gamma (V)}-\gamma (V)$, we are able to find
\begin{equation}\label{56}
\left\{\begin{array}{l}
 \gamma (V)=\frac{1}{\sqrt{1-\frac{V^2}{C_R^2}}}\\
 \alpha (V)=-\frac{V}{C_R^2}\frac{1}{\sqrt{1-\frac{V^2}{C_R^2}}}.
\end{array}\right.
\end{equation}
Apparently, relations $\gamma (-V)=\gamma (V)$ and $\alpha(-V)=-\alpha (V)$, as well as conditions $\gamma (V=0)=1$ and $\alpha (V=0)=0$ are satisfied. Besides, $-\frac{\alpha (V)}{\gamma (V)}$ and $V$ have the same signs, so that invariant velocity exists. On the other hand, from the condition $1+\frac{\alpha (V)}{\gamma (V)}V=1-\frac{V^2}{C_R^2}>0 $, it is guaranteed that the condition $|V|<C_R$ must be satisfied for relative velocities between inertial frames. Now consider one particle $m_{01}>0$ with speed limit $v_{max1}$ and another particle $m_{02}>0$  with speed limit $v_{max2}$ . Given the relative velocity of $V$ between them, in either of the rest frames, the other particle in principle can approach the limit speed $v_1=|\vec{v}_1|=|V|\rightarrow v_{max1}$, $v_2=|\vec{v}_2|=|V|\rightarrow v_{max2}$. On the other hand, $V$ being the relative speed between frames must be subject to $|V|<v_{max1}$ ,  $|V|<v_{max2}$ and $|V|<C_R$. It can therefore be deduced that all particle with $m_0>0$ are subject to the same universal speed limit $v_{max1}=v_{max2}=C_R$.

Substituting $\gamma (V)$ and $\alpha (V)$ into Eq.~(\ref{1.2}), we obtain the relativistic transformation rules for space and time coordinates
\begin{equation}{\label{57}}
\left\{\begin{array}{l}
x'=\frac {x-V t}{\sqrt{1-\frac{V^2}{C_R^2}}}\\
y'=y\\
z'=z\\
t'=\frac {t-\frac{V}{C_R^2}x}{\sqrt{1-\frac{V^2}{C_R^2}}}.
\end{array}\right.
~~~~(|V|<C_R)
\end{equation}
We dub them as {\it generalized Lorentz transformation}, with the Relativity Constant $C_R$ replacing the speed of light $c$ in the familiar Lorentz transformations. From Eq.~(\ref{1.8}), relativistic velocity transformation read
\begin{equation}{\label{58}}
\left\{\begin{array}{l}
v'_x= \frac{v_x-V}{1-(V v_x)/C_R^2}\\
\\
v'_y=\frac{v_y \sqrt{1-{V^2}/C_R^2}}{1-(V v_x)/C_R^2}\\
\\
v'_z=\frac{v_z \sqrt{1-{V^2}/C_R^2}}{1-(V v_x)/C_R^2}.
\end{array}\right.
~~~~(|V|<C_R)
\end{equation}
Let $v_{\parallel}=v_x$, $v_{\perp}^2=v_y^2+v_z^2=v^2-v_{\parallel}^2$, and after some algebra we obtain
\begin{equation}\label{59}
v'^2=\frac{(V-v_x)^2+(1-\frac{V^2}{C_R^2})(v_y^2+v_z^2)}{(1-\frac{V}{C_R^2}v_x)^2}
=\frac{C_R^2(V- v_\parallel)^2+(C_R^2-V^2)v_{\perp}^2}{(C_R-\frac{V}{C_R}v_{\parallel})^2}.
\end{equation}
It is clear that either $v_{\perp}=0$, $v_{\parallel}\rightarrow C_R$ or $v_{\parallel}=0$, $v_{\perp}\rightarrow C_R$  leads to particle velocity $v\rightarrow C_R$ and $v'\rightarrow C_R$ simultaneously in both frames. Besides, in the case of $|V|\rightarrow C_R$, the above equation also implies $v'\rightarrow C_R$. This validates the previous claim that the Relativity Constant $C_R$ is nothing but the invariant speed under the generalized Lorentz transformations between inertial frames. If we take $C_R=c$, the generalized Lorentz transformation will reduce to conventional Lorentz transformation. Then in our new framework, the constant speed of light arises as a derived rather than assumed quantity. The generalized Lorentz transformation reduces to the classical Galilean transformation in the limit $C_R\rightarrow \infty$, so that the Galilean relativity with $\gamma (V)=1$ can be included as a special limit $C_R\rightarrow \infty$ for the generalized Lorentz transformation~\cite{16,17}.

\subsection{\bf Particles having no rest frames}

In the above sub-Sections, by scrutinizing massive particles with a notion of rest frame, we have been conforming to the principle of relativity and have based our derivations on basic dynamic definitions $\vec{p}= m\vec{v}$, $d\vec{p}= \vec{F}dt$, $dE=\vec{F}\cdot d\vec{r}$, as well as the conservation laws for momentum and energy, and have derived the relativistic mass-velocity and mass-energy relations, without making use of any particular transformation rules of space-time. Eventually, this enables us to determine the transformation rules for space-time, i.e. the generalized Lorentz transformation.

Even for a particle that does not find a rest frame, it conforms to the same relativistic dynamics, i.e. $\vec{p}= m\vec{v}$, $d\vec{p}= \vec{F}dt$, $dE=\vec{F}\cdot d\vec{r}$, where $m=m(v^2)$, $E=E(v^2)$, and its energy is also expected to be proportional to its moving mass, i.e. $E(v^2)=m(v^2)B$ with $B$ being a constant. Particularly, the corresponding mass-velocity relation should as well solve the differential equation of Eq.~(\ref{49})
\begin{equation}\label{60}
\frac{dm}{m}=\frac{d(v^2)}{2(B-v^2)}.
\end{equation}

A different solution exists for the same different equation: $\ln m(v^2)=-\frac{1}{2}\ln(v^2-B)+C$, where again $C$ is a constant of integration. From the logarithmic term it is required that $v^2-B>0 $, namely
\begin{equation}\label{61}
m=m(v^2)=\frac{e^{C}}{\sqrt{v^2-B}}~~~~~(v^2>B).
\end{equation}
The condition $v^2>B$ suggests that a lower bound for speed exists for particles having no rest frames. Let $v_{min}$ be such a lower bound, which can also be saturated. Let $V>0$, taking $\pm V$ both in the velocity transforms formula, in the limit $|v|\rightarrow v_{min}$, we have
\begin{equation}\label{62}
\left\{\begin{array}{l}
 |v'|=|\frac{v_{min}-V}{1-\frac{V}{C_R^2}v_{min}}|\geq v_{min}\\
 |v'|=|\frac{v_{min}+V}{1+\frac{V}{C_R^2}v_{min}}|\geq v_{min}.
\end{array}\right.
~~~
\left\{\begin{array}{l}
 |v_{min}-V|\geq |v_{min}-\frac{v_{min}^2}{C_R^2}V|\\
 |v_{min}+V|\geq |v_{min}+\frac{v_{min}^2}{C_R^2}V|.
\end{array}\right.
\end{equation}
i.e.
\begin{equation}\label{63}
\left\{\begin{array}{l}
 -|v_{min}-V|\leq (v_{min}-\frac{v_{min}^2}{C_R^2}V)\leq |v_{min}-V|\\
 -|v_{min}+V|\leq (v_{min}+\frac{v_{min}^2}{C_R^2}V)\leq |v_{min}+V|.
\end{array}\right.
\end{equation}
Since $|v|\geq v_{min}$, we have $V \geq v_{min}$ is forbidden by virtue of the fact that no rest frame exists, it is suggested that for $V <v_{min}$ we have inequalities
\begin{equation}\label{64}
\left\{\begin{array}{l}
 ~(V- v_{min})\leq (v_{min}-\frac{v_{min}^2}{C_R^2}V)\leq (v_{min}-V)\\
 -(v_{min}+V)\leq (v_{min}+\frac{v_{min}^2}{C_R^2}V)\leq (v_{min}+V).
\end{array}\right.
\end{equation}
This can be turned into
\begin{equation}\label{65}
\left\{\begin{array}{l}
 ~~(1+\frac{v_{min}^2}{C_R^2})V\leq 2v_{min}~~~,~~~v_{min}^2\geq C_R^2\\
 -(1+\frac{v_{min}^2}{C_R^2})V\leq 2v_{min}~~~,~~~v_{min}^2\leq C_R^2.
\end{array}\right.
\end{equation}
It can be inferred from the above equation that $|V|<v_{min}=C_R$, which proves that the lower velocity bound for those particle species without rest frames must in terms of its value coincide with the Relativity Constant
\begin{equation}\label{66}
 v_{min}=C_R.
\end{equation}
Now let $v=|\vec{v}|$, $p=|\vec{p}|=mv$, then external force $\vec{F}$ projects onto the tangent direction of particle motion through
\begin{equation}\label{67}
 F_{t}=\frac{dp}{dt}=m\frac{dv}{dt}+v\frac{dm}{dt}=\frac{dv}{dt}(m+v\frac{dm}{dv})=a_t\frac{-e^{C}B}{(v^2-B)^{\frac{3}{2}}}.
\end{equation}
Acted on by the external force, the particle decelerates to approach the lower bound $C_R$. In the meantime, the tangent component of its acceleration must also diminish $\lim a_t=\lim \frac{dv}{dt}=0$ as $v \rightarrow C_R$. If some non-zero force is still applied in the tangent direction to the particle, from Eq.~(\ref{67}) it has to be that $\lim_{v \rightarrow C_R}~ (v^2-B)^{\frac{3}{2}}\rightarrow 0$ , from which we determine $B=C_R^2$. We point out that using the same method one can also determine $A=C_R^2$ in Eq.~(\ref{50}). For the case of $\vec {v}=\vec {v_0}$, the particle has mass $m(v_0^2)=\frac{e^{C}}{\sqrt{v_0^2-B}}$, i.e. $e^{C}=m(v_0^2)\sqrt{v_0^2-B}$. We therefore establish the mass-velocity relation for such particles having no rest frames
\begin{equation}\label{68}
 m=m(v^2)=m(v_0^2)\sqrt{\frac{v_0^2-C_R^2}{v^2-C_R^2}}~~~~(|\vec {v}|>C_R ).
\end{equation}
The mass-energy relation then naturally follows
\begin{equation}\label{69}
E=E(v^2)=m(v^2)C_R^2~~~~(|\vec {v}|>C_R ).
\end{equation}
If one sets $\vec {v}=0$, then $m_0=m(0)$ is the equivalent of the rest mass. The mass-velocity relation for imaginary rest mass tachyon reads
\begin{equation}\label{70}
 m=\frac{m_0}{\sqrt{1-\frac{v^2}{C_R^2}}}~~~~(|\vec {v}|>C_R ).
\end{equation}
This corresponds to the energy for imaginary rest mass tachyon
\begin{equation}\label{71}
 E=\frac{m_0C_R^2}{\sqrt{1-\frac{v^2}{C_R^2}}}~~~~(|\vec {v}|>C_R ).
\end{equation}
Since energy must be a real number, the rest mass $m_0$  has to be purely imaginary~\cite{11,18}.

It is conventionally believed that tachyons do not define any rest frames and they travel faster than the Relativity Constant: $|\vec {v}|>C_R$. There is  no consistent and meaningful definition of a tachyonic rest mass. Neither is it found in the literature, nor is it plausible to measure it in experiments even if any tachyon is observed in the real world. We believe $m(v_0^2)$ in Eq.~(\ref{68}) should be undertood as the mass for particle species which can not define rest frames and which belong to the case of $|\vec {v}|>C_R$. From that we obtain
\begin{equation}\label{72}
 m(v_0^2)\sqrt{v_0^2-C_R^2}=m\sqrt{v^2-C_R^2}=\sqrt{\vec {p}^2-m^2C_R^2}.
\end{equation}
Since $m_{\infty}=m(v^2\rightarrow \infty)\rightarrow 0$, let $\vec {p}_{\infty}$ be the momentum of tachyon at $\vec {v}\rightarrow \infty$. Then we have $m(v_0^2)\sqrt{v_0^2-C_R^2}=|\vec {p}_{\infty}|$, so the mass-velocity relation for a real-mass tachyon reads
\begin{equation}\label{73}
 m=\frac{|\vec {p}_{\infty}|}{\sqrt{v^2-C_R^2}}~~~~(|\vec {v}|>C_R ).
\end{equation}
The energy for a real-mass tachyon is
\begin{equation}\label{74}
 E=mC_R^2=\frac{|\vec {p}_{\infty}|C_R^2}{\sqrt{v^2-C_R^2}}~~~~(|\vec {v}|>C_R ).
\end{equation}

For particle moving along the $X$ direction, we introduce dimensionless (subscript $x$ dropped) ratios $\bar{v}'=\frac{v'}{C_R}$, $\bar{v}=\frac{v}{C_R}$, $\bar{V}=\frac{V}{C_R}$, so the velocity transformation becomes
\begin{equation}\label{75}
\bar{v}'=f(\bar{v}, \bar{V})=\frac{\bar{v}-\bar{V}}{1-\bar{v} \bar{V}}~~~~(|\bar {v}|<1).
\end{equation}

\begin{figure}[ht]
  \includegraphics[scale=0.4]{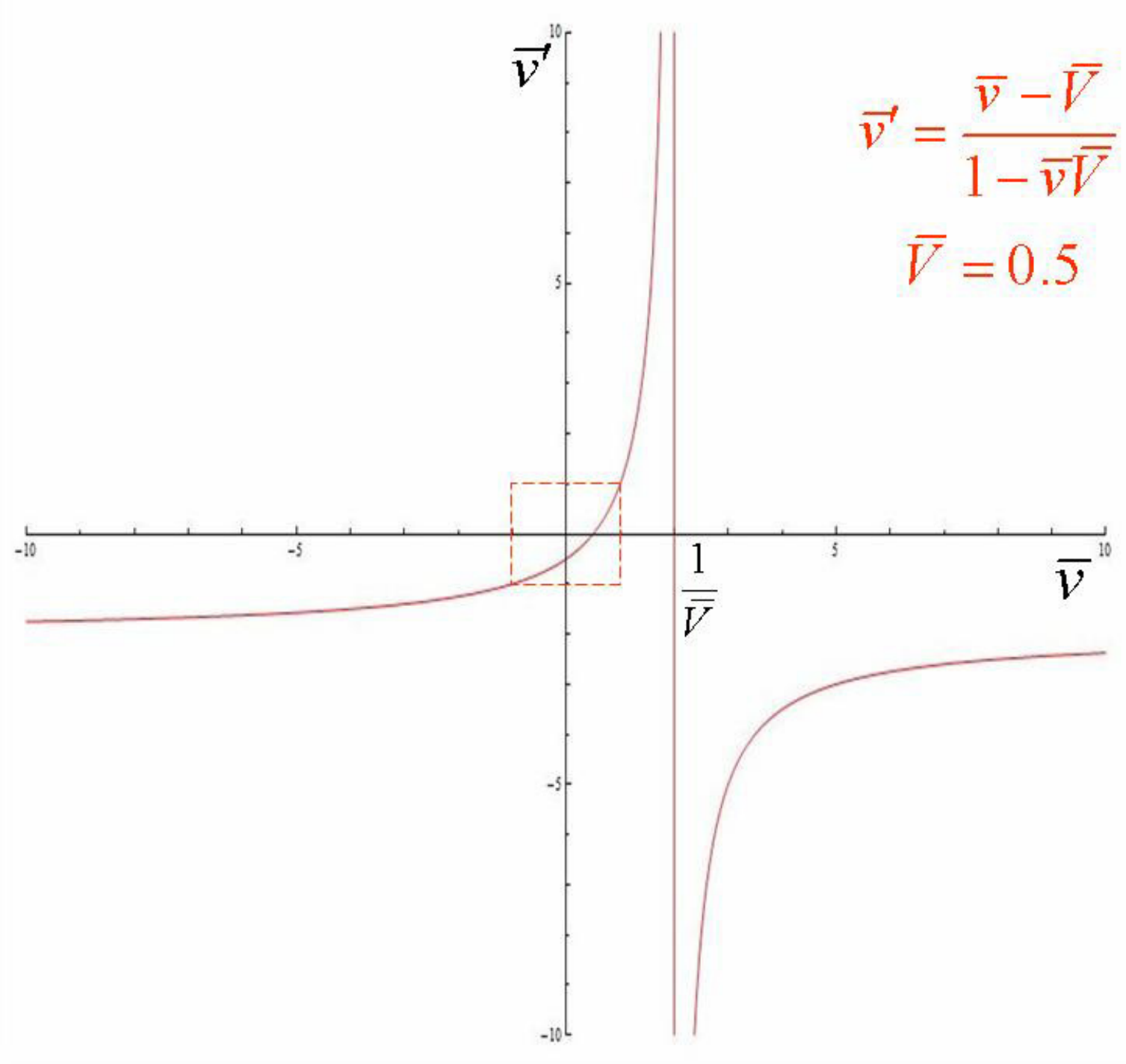}\\
  \caption{The boost of dimensionless generalized Lorentz transformation with $\bar{V}=0.5$} \label{fig:8}
\end{figure}

\begin{figure}[ht]
  \includegraphics[scale=0.4]{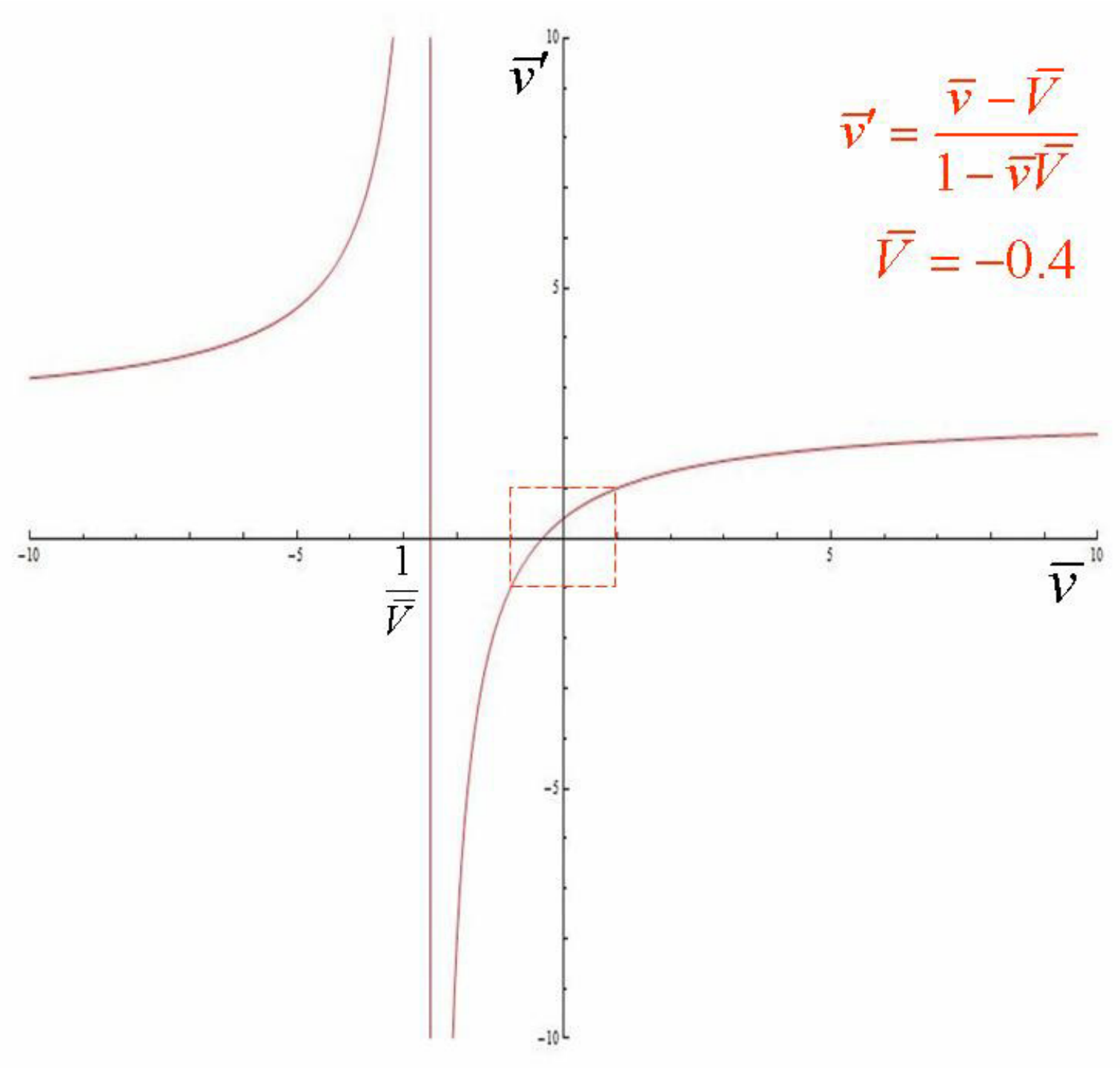}\\
  \caption{The boost of dimensionless generalized Lorentz transformation with $\bar{V}=-0.4$} \label{fig:9}
\end{figure}

It can be verified that $|\bar{v}|=1$ , (i.e. $|v|=C_R $), is a fixed point for the velocity transformation. For tardyons with non-zero rest mass, $-1<\bar{v}<1 $ and $\bar{v}'=f(\bar{v}, \bar{V})$ is continuous, monotonically increasing and bounded from above $|\bar{v}|<1$ . As for tachyons with lower bound $|\bar{v}|>1$, $\bar{v}\in (-\infty, -1)$ or $\bar{v}\in (1, \infty)$, the transformed velocity $|\bar{v}'|>1$ , and $\bar{v}'=f(\bar{v}, \bar{V})\rightarrow \pm \infty$ at the discontinuity $\bar{v}=\frac{1}{\bar{V}}$. Therefore, a tachyon with original velocity $\bar{v}$ will have infinite velocity after a frame boost $\bar{V}=\frac{1}{\bar{v}}$, and the value for momentum of the tachyon $|\vec {p}_{\infty}|$ can be determined. The generalized Lorentz transformation parameterized by the dimensionless boost velocity for $\bar{V}=0.5$ and for $\bar{V}=-0.4$ are shown in Figure~\ref{fig:8} and Figure~\ref{fig:9} respectively. Tardyons have non-zero rest mass with $|\bar{v}|<1$, shown as the dashed line in the Figures, while the tachyons have $|\bar{v}'|>1$. Given that no tachyon has been observed in experiments, the particle's intrinsic momentum parameter $|\vec {p}_{\infty}|$ is yet to be known. We propose that the correct tachyon mass-velocity relation is Eq.~(\ref{73}), where the real-mass tachyon may be a candidate for the Dark Matter~\cite{19,20}.


\section{Relativistic transformation for energy-momentum}
\label{sec: energy-momentum-transform}

From previous discussions, the relativistic momentum and energy are $\vec{p}= m\vec{v}$, $E= mC_R^2$ respectively for all particles. Because the mass-velocity relation is not same for tardyons and constons, so there are different energy-momentum relations for them. But for all particles, the quadratic form of energy and momentum $ E^2-\vec{p}^2C_R^2$ is invariant under transformation between different inertial frames, from which the generalized Lorentz transformation can also be derived.

\subsection{\bf Relativistic energy-momentum relation}

In this sub-Section, we examine below several possibilities of relativistic energy-momentum relations depending on whether the particle defines a rest frame:

(i)Tardyons having rest frames

The mass-velocity relation is $m=m_0[1-(v^2/C_R^2)]^{-\frac{1}{2}}$ with $|\vec{v}| < C_R$, where $m_0$ is the intrinsic mass or rest mass for tardyons. Using $\vec{p}= m\vec{v}$, $E= mC_R^2$, there is $\frac{\vec{p}^2}{E^2}=\frac{v^2}{C_R^4}$, we eliminate $v^2$ and obtain the energy-momentum relation for tardyons
\begin{equation}\label{76}
E^2=\vec{p}^2C_R^2- m_0^2C_R^4.
\end{equation}

(ii)Tachyons having no rest frames

The mass-velocity relation reads $m=|\vec{p}_\infty |(v^2-C_R^2)^{-\frac{1}{2}}$ with $|\vec{v}| > C_R$, where $|\vec{p}_\infty|$ is the momentum for tachyon in the limit $|\vec{v}|\rightarrow \infty$. From $\vec{p}= m\vec{v}$, we have
\begin{equation}\label{77}
\vec{p}=\frac{|\vec{p}_\infty|\vec{v}}{\sqrt {v^2-C_R^2}} ~~~~ (|\vec{v}| > C_R).
\end{equation}
The $|\vec{p}_\infty|$ can be thought of as the intrinsic momentum for a real-mass tachyon. The above equation establishes the relation between the tachyonic momentum and the intrinsic momentum, which can be written as
\begin{equation}\label{78}
\vec{p}^2=m^2v^2=\frac{{\vec{p}_\infty}^2}{1-\frac{C_R^2}{v^2}}\geq{\vec{p}_\infty}^2.
\end{equation}
Namely, $|\vec{p}_\infty|$ is the minimum momentum for tachyon. The energy-momentum relation for tachyon is then obtained by eliminating $v^2$ from $E=mC_R^2=\frac{|\vec{p}_\infty|C_R^2}{\sqrt {v^2-C_R^2}}$ and $\frac{\vec{p}^2}{E^2}=\frac{v^2}{C_R^4}$. It reads
\begin{equation}\label{79}
E^2=\vec{p}^2C_R^2-{\vec{p}_\infty}^2C_R^2.
\end{equation}
The intrinsic momentum can be determined from the momentum and energy for tachyons by relation $|\vec{p}_\infty|=\sqrt {\vec{p}^2-\frac{E^2}{C_R^2}}$.

(iii)Constons with invariant speed and without rest frames

For constant-speed particles which we call {\it constons}, the momentum and energy are $|\vec{p}|= mC_R$, $E= mC_R^2$ respectively, and the energy-momentum relation is obtained by eliminating $m$
\begin{equation}\label{80}
E^2=\vec{p}^2C_R^2.
\end{equation}

From the above discussion, there is an intrinsic mass or minimum mass $m_0$ for tardyons, intrinsic momentum or minimum momentum $|\vec{p}_\infty|$ for tachyons and intrinsic speed or constant-speed $C_R$ for constons, respectively.
The complete energy-momentum relation for all particles reads
\begin{equation}\label{81}
E^2-\vec{p}^2C_R^2
\left\{\begin{array}{l}
 =m_0^2C_R^4>0~~~~~~~~( tardyon:~ |\vec{v}| < C_R) \\
 =0~~~~~~~~~~~~~~~~~~~~( conston:~ |\vec{v}| = C_R) \\
 =-{\vec{p}_\infty}^2C_R^2 <0~~~~~( tachyon:~ |\vec{v}| > C_R)
\end{array}\right.
\end{equation}

It should be pointed out that neither tardyons accelerating to the upper speed limit: $|\vec{v}|\rightarrow C_R$, nor tachyons decelerating to the lower speed limit: $C_R \leftarrow|\vec{v}|$, can be identified with constons. Therefore, the three categories of particles are intrinsically different from each other, according to the principle of relativity, which also requires that $E^2-\vec{p}^2C_R^2$ is an invariant quantity under transformations between reference frames.

\subsection{\bf Relativistic transformation for energy-momentum}

\begin{figure}[ht]
  \includegraphics[scale=0.4]{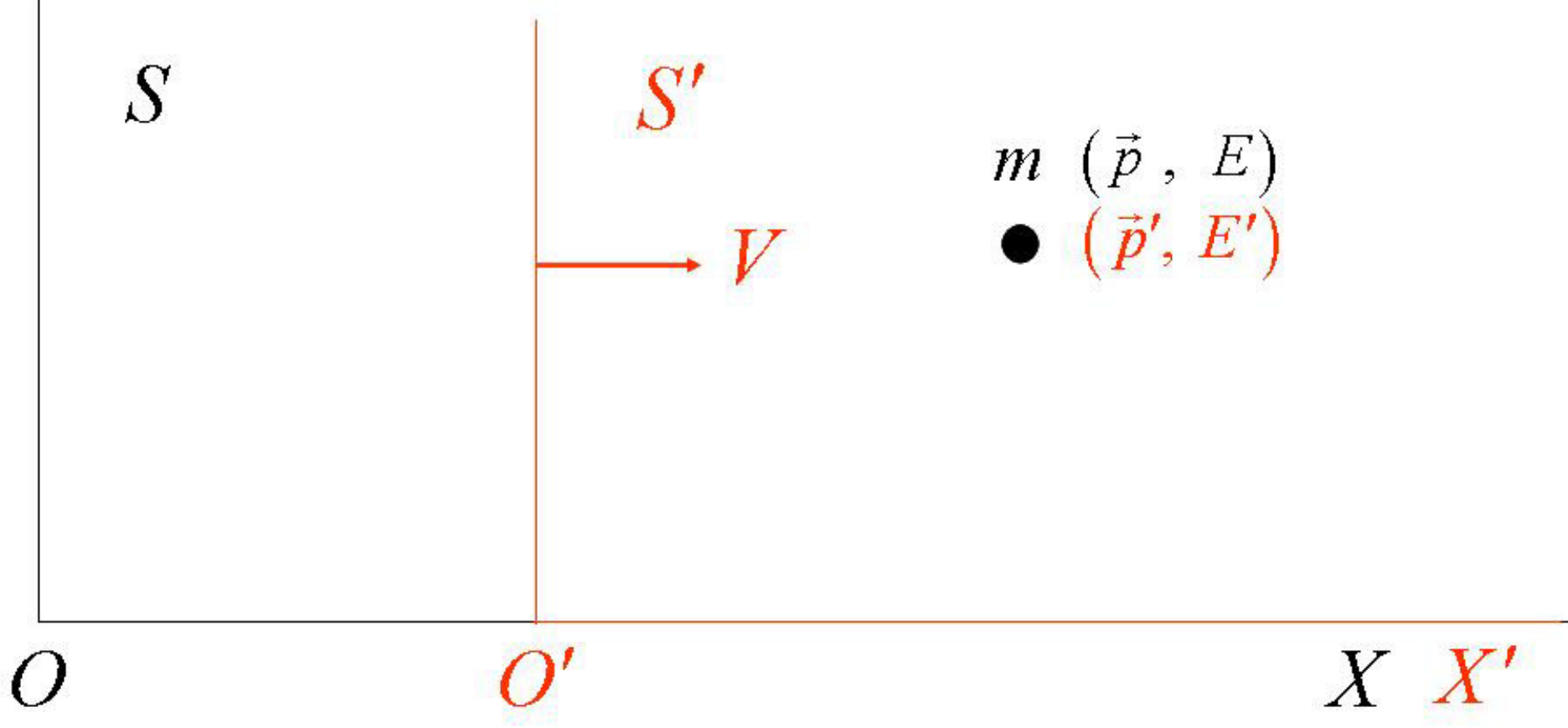}\\
  \caption{Transformation of energy-momentum in different inertial frames} \label{fig:10}
\end{figure}

Consider a particle that has a rest frame (i.e. a tardyon) is at rest in inertial frame $S'$ , and this frame has relative velocity $V$ to another inertial frame $S$ along the $X$ direction, and hence the particle has velocity $|\vec{v}|=V$ as observed in frame $S$. Since the tardyonic velocity satisfies $|\vec{v}| < C_R$, we restrict the inertial frame boost to $|V| < C_R$. In frame $S$ and frame $S'$, the particle has momentum-energy $(\vec{p},~ E)$ and $(\vec{p}',~ E')$, respectively, as shown in Figure~\ref{fig:10}. Since both particle's momentum $\vec{p}= m_0\gamma \vec{v}$ and energy $E= m_0\gamma C_R^2$  are proportional to the intrinsic mass $m_0$, in order to have additive $m_0$ they should transform linearly between different inertial frames. Most generally, we can write

\begin{equation}\label{82}
\left\{\begin{array}{l}
 p'_x=a_1p_x+ a_2E \\
 p'_y=p_y\\
 p'_z=p_z\\
 E'=a_3p_x+ a_4E.
\end{array}\right.
\end{equation}
Following that in frame $S'$ particle's velocity $\vec{v}=0$, while in frame $S$ the velocity is $v_x=V$, we have $0=a_1mV+ a_2mC_R^2$ so that $a_2=-a_1\frac{V}{C_R^2}$, and $m_0C_R^2=a_3mV+ a_4mC_R^2$ so that $a_4=\sqrt{1-\frac{V^2}{C_R^2}}-a_3\frac{V}{C_R^2}$. The transformation reads

\begin{equation}\label{83}
\left\{\begin{array}{l}
 p'_x=a_1(p_x- \frac{V}{C_R^2}E)\\
 p'_y=p_y\\
 p'_z=p_z\\
 E'=a_3(p_x-\frac{V}{C_R^2}E)+ \sqrt{1-\frac{V^2}{C_R^2}}E.
\end{array}\right.
\end{equation}
According to the foregoing discussion, there is the invariant quantity $E'^2-\vec{p}'^2C_R^2= E^2-\vec{p}^2C_R^2$ under transformation of frames, so we obtain

\begin{equation}\label{84}
\left\{\begin{array}{l}
\{(a_3^2-a_1^2C_R^2)\frac{V^2}{C_R^4}-2a_3\frac{V}{C_R^2}\sqrt{1-\frac{V^2}{C_R^2}}-\frac{V^2}{C_R^2}+1\}E^2=E^2\\
(a_3^2-a_1^2C_R^2)p_x^2=-C_R^2p_x^2\\
\{2a_3\sqrt{1-\frac{V^2}{C_R^2}}-2(a_3^2-a_1^2C_R^2)\frac{V}{C_R^2}\}{p_x}E=0.
\end{array}\right.
\end{equation}
Note that only two of equations are independent. Then we are able to determine $a_1$ and $a_3$ from above relations

\begin{equation}\label{85}
\left\{\begin{array}{l}
a_1=\frac{1}{\sqrt{1-\frac{V^2}{C_R^2}}}\\
a_3=\frac{-V}{\sqrt{1-\frac{V^2}{C_R^2}}}.
\end{array}\right.
\end{equation}
Inserting $a_1$ and $a_3$ into Eq.~(\ref{83}), we obtain

\begin{equation}\label{86}
\left\{\begin{array}{l}
 p'_x=\frac{p_x- \frac{V}{C_R^2}E}{\sqrt{1-\frac{V^2}{C_R^2}}}\\
 p'_y=p_y\\
 p'_z=p_z\\
 E'=\frac{E-Vp_x}{\sqrt{1-\frac{V^2}{C_R^2}}}.
\end{array}\right.
~~~~~~(|V|<C_R)
\end{equation}
This is nothing but the familiar Lorentz transformation for energy-momentum with the Relativity Constant $C_R$ replacing the speed of light $c$~\cite{17}.

The above derivation for the energy-momentum transformation is aimed at tardyons. However, one can verify directly that tachyons and constons also satisfy $E'^2-\vec{p}'^2C_R^2= E^2-\vec{p}^2C_R^2$ as well. From that transformation relation, we have

\begin{equation}\label{87}
m'v'_x=\frac{mv_x- \frac{V}{C_R^2}mC_R^2}{\sqrt{1-\frac{V^2}{C_R^2}}}=\frac{m(v_x- V)}{\sqrt{1-\frac{V^2}{C_R^2}}}.
\end{equation}
This means
\begin{equation}\label{88}
\frac{{v'_x}^2}{1-\frac{{v'_x}^2}{C_R^2}}=\frac{(v_x-V)^2}{(1-\frac{{v_x}^2}{C_R^2})(1-\frac{V^2}{C_R^2})}.
\end{equation}
We solve for the transformation relation of velocity

\begin{equation}\label{89}
v'_x=\frac{v_x-V}{1-\frac{V}{C_R^2}{v_x}}.
\end{equation}
Comparing this with the general form for velocity transformation in Eq.~(\ref{10}), we can obtain $\frac{\alpha (V)}{\gamma (V)}=-\frac{V}{C_R^2}$. Once again, we can derive the generalized Lorentz transformation relying not on any kinematical assumption, but on the relativistic invariant quantity $E^2-\vec{p}^2C_R^2$, the relation $\alpha (V)V=\frac{1}{\gamma (V)}-\gamma (V)$ and the condition $\gamma (V=0)=1$. It is specified by the following two transformation functions

\begin{equation}\label{90}
\left\{\begin{array}{l}
 \gamma (V)=\frac{1}{\sqrt{1-\frac{V^2}{C_R^2}}}\\
 \alpha (V)=-\frac{V}{C_R^2}\frac{1}{\sqrt{1-\frac{V^2}{C_R^2}}}.
\end{array}\right.
\end{equation}
Denote the four-dimensional coordinate $( x_1, x_2, x_3, x_4)=( x, y, z,{C_R}t)$, and the four-dimensional momentum $( p_1, p_2, p_3, p_4)=( p_x, p_y, p_z, \frac{E}{C_R})$, the four-dimensional form for the generalized Lorentz transformation is obtained
\begin{equation}\label{91}
\left(
  \begin{array}{c}
    x'_1 \\
    x'_2 \\
    x'_3 \\
    x'_4 \\
  \end{array}
\right)
=
\left(
   \begin{array}{cccc}
     \gamma & 0 & 0 & -\frac{V}{C_R}\gamma \\
     0 & 1 & 0 & 0 \\
     0 & 0 & 1 & 0 \\
     -\frac{V}{C_R}\gamma & 0 & 0 & \gamma \\
   \end{array}
 \right)
 \left(
   \begin{array}{c}
     x_1 \\
     x_2 \\
     x_3 \\
     x_4 \\
   \end{array}
 \right).
 ~~~~~~(|V|<C_R)
\end{equation}

\begin{equation}\label{92}
\left(
  \begin{array}{c}
    p'_1 \\
    p'_2 \\
    p'_3 \\
    p'_4 \\
  \end{array}
\right)
=
\left(
   \begin{array}{cccc}
     \gamma & 0 & 0 & -\frac{V}{C_R}\gamma \\
     0 & 1 & 0 & 0 \\
     0 & 0 & 1 & 0 \\
     -\frac{V}{C_R}\gamma & 0 & 0 & \gamma \\
   \end{array}
 \right)
 \left(
   \begin{array}{c}
     p_1 \\
     p_2 \\
     p_3 \\
     p_4 \\
   \end{array}
 \right).
 ~~~~~~(|V|<C_R)
\end{equation}
Then we have
\begin{equation}\label{93}
p^2=p_1^2+ p_2^2+ p_3^2- p_4^2=\vec{p}^2- \frac{E^2}{C_R^2}
\left\{\begin{array}{l}
 =-\frac{E_0^2}{C_R^2}<0~~~~~~~~( tardyon:~ |\vec{v}|< C_R) \\
 =0~~~~~~~~~~~~~~~~~~~( conston:~ |\vec{v}| = C_R) \\
 =|\vec{p}_\infty|^2>0.~~~~~~~( tachyon:~ |\vec{v}|> C_R)\\
\end{array}\right.
\end{equation}
In the case of $\vec{v}=0$ for tardyons, we have $m(0)=m_0$, $\vec{p}=m\vec{v}=0$, $E=E_0=m_0C_R^2 \neq0$, where $E_0$ is the rest energy (or minimum energy) of the tardyon. By comparison, In the case of $|\vec{v}|\rightarrow \infty$ for tachyons, we have $m(\infty)=0$, $E=mC_R^2=0$, $\vec{p}=\vec{p}_\infty \neq0$, where $|\vec{p}_\infty|$ is the intrinsic momentum (or minimum momentum) of the tachyon. The quadratic form of the four-dimensiononl momentum, $p^2 <0$ for tardyons, $p^2 =0$ for constons and $p^2 >0$ for tachyons respectively, is invariant.


\section{Conclusion and discussion}
\label{sec:conclusion}

In this paper, we have presented a novel derivation of Special Relativity for the full speed range, which does not rely on the assumption of constant speed of light. The discussion consists of the following crucial parts:

(1) For kinematics, we have proved that for particles which have rest frames, the Galilean transformation with absolute notion of time is the only consistent linear transformation of space-time that allows arbitrarily large velocity of motion. Therefore, for non-Galilean transformation, upper speed bound has to exist for particles having rest frame. On the other hand, we have also shown that particles without rest frame or rest mass, are subject to a lower speed bound instead.\\

(2) For dynamics, from the principle of relativity and conservation of relativistic momentum and energy, we have derived without utilizing any particular form of space-time transformation that for massive particle species the relativistic energy has to be proportional to the relativistic mass. We also derived a differential equation for the mass-velocity relation, and obtained two distinct solutions. One of the solutions $m=m_0[1-(v^2/C_R^2)]^{-\frac{1}{2}}$ applies to tardyons which have rest frames, and suggests an upper bound for speed $|\vec{v}| < C_R$. The other solution $m=|\vec{p}_\infty |(v^2-C_R^2)^{-\frac{1}{2}}$ with $ \vec{p}_\infty $ being the finite momentum at infinite speed, describes tachyons which have no rest frames, and suggests lower bound for speed $|\vec{v}| > C_R$. The universal constant $C_R$, what we call the Relativity Constant, is required by relativity. Given the corresponding generalized Lorentz transformation $\gamma(V)=[1-(V^2/C_R^2)]^{-\frac{1}{2}}$ with $|V| < C_R$, it means that $C_R$ also is an invariant velocity for the case of $\gamma(V)>1 $. The derivation implies that the Relativity Constant $C_R$ determines the property of space-time. The infinite limit $C_R\rightarrow \infty$ is equivalent to a Galilean transformation with $\gamma(V)=1 $. The usual Lorentz transformation arises when the Relativity Constant equal to the speed of light $C_R=c$, and consequently the constant speed of light becomes derived rather than assumed. A most general value for $C_R$ corresponds to the general Lorentz transformation. The case of Galilean transformation has been ruled out because experiments indicate that upper bound of particle velocity does exist for tardyons. Another way to look at the limit $C_R\rightarrow \infty$ case is that the rest energy will diverge $E_0=m_0C_R^2\rightarrow \infty$, and so does the energy $E=mC_R^2\rightarrow \infty$, which is physically unreasonable. As a matter of fact, the limit $C_R\rightarrow \infty$  only applies in the classical Newtonian picture, with momentum $\vec p\rightarrow m_0\vec{v}$ and kinetic energy $E_{k}=E-E_0 \rightarrow \frac{1}{2}m_{0}v^2$, and otherwise contradicts with the relativistic picture. Hence a finite $C_R$ is physical, whose precise value can be determined by experiment.\\

(3) The relativistic momentum and energy are $\vec{p}= m\vec{v}$, $E= mC_R^2$ for all particles. However, particles from different categories have different mass-velocity relations. For tardyons which satisfy $|\vec{v}| < C_R$ and have rest frames and intrinsic mass, namely the rest mass, the energy-momentum relation reads $E^2-\vec{p}^2C_R^2 =m_0^2C_R^4 >0$. For tachyons which satisfy $|\vec{v}| > C_R$ and have no rest mass but with intrinsic momentum $|\vec{p}_\infty| $, its energy-momentum relation reads $E^2-\vec{p}^2C_R^2 =-\vec{p}_\infty^2 C_R^2<0$. For constons which move at invariant speed $|\vec{v}| = C_R$, the notion of rest frame or rest mass is also meaningless, but the momentum and energy are given by $|\vec{p}| =mC_R$, $E= mC_R^2$ respectively, and the corresponding energy-momentum relation reads $E^2-\vec{p}^2C_R^2 =0$. Therefore, for the full velocity range $|\vec{v}| < C_R$ and $|\vec{v}| \geq C_R$ , the quantity $E^2-\vec{p}^2C_R^2$ becomes invariant among all frames. In the traditional formulation of relativity, the relativistic dynamics are derived from the Lorentz transformation, which is based on the assumption of constant speed of light, the principle of relativity as well as basic dynamic definitions $\vec{p}= m\vec{v}$, $d\vec{p}= \vec{F}dt$, $dE=\vec{F}\cdot d\vec{r}$. We have shown in this work that the assumption of constant speed of light is redundant. In other words, relativistic dynamics can be derived from solely the principle of relativity and basic definitions $\vec{p}= m\vec{v}$, $d\vec{p}= \vec{F}dt$, $dE=\vec{F}\cdot d\vec{r}$. Then the generalized Lorentz transformation of space-time is uniquely fixed from the invariant quantity $E^2-\vec{p}^2C_R^2$ under transformation between frames.

In conventional relativity, the kinematic quantity $s^2=\vec{r}^2-c^2t^2$ is invariant under Lorentz transformation, with the speed of light $c$ being a universal constant introduced by the assumption of constant speed of light. From that, the mass-velocity and mass-energy relations for relativistic dynamics can be obtained. Here $s^2>0$, $s^2 =0$ and $s^2<0$ corresponds to space-like, light-like and time-like separations, respectively. Conversely, our relativistic formalism is based on the dynamical quantity $p^2=\vec{p}^2-(E^2/C_R^2)$, invariant in all inertial frames. And the Relativity Constant $C_R$ is required by the principle of relativity. Therefore, the generalized Lorentz transformation can be determined without the assumption of constant speed of light, and the three cases $p^2=-m_0^2 C_R^2<0$, $p^2=0$ and $p^2=|\vec{p}_\infty|^2>0$ describe tardyons, constons and tachyons respectively. In contrast to the conventional relativistic dynamics which are based on relativistic kinematics, we take an alternative route, i.e. determine relativistic kinematics from relativistic dynamics.\\

(4) In traditional formulation of relativity, the energy-momentum relation is $E^2-\vec{p}^2c^2 =m_0^2c^4 >0$, and the photon has zero rest mass $m_{0}=0$, while the tachyon has rest mass $m_{0}^2<0$, which has to be imaginary rest mass for tachyons~\cite{11,21}. From our point of view in this work, the notion of rest mass is only meaningful for tardyons with rest frames such that $|\vec{v}| < C_R$ and $m_{0}^2 >0$. Therefore, there should be no constons with zero rest mass, nor is the imaginary rest mass of tachyons meaningful. That is, no particle species with $m_{0}^2 \leq0$ exist in Nature. If photons move at invariant speed $|\vec{v}|= C_R$ and belong to the category of constons, the notion of rest mass or rest energy should not be introduced for them. Indeed, in experiments it is not possible to measure the rest mass and the rest energy of a photon. Nonetheless, its momentum and energy can be measured, and the condition $E^2-\vec{p}^2C_R^2=0$  can then be tested to judge whether or not it is a conston. For tachyons which have lower speed limit, no rest frame or rest mass can be defined either. Different from the usual description in terms of an imaginary rest mass, tachyons have intrinsic momentum (or minimal momentum) and real mass at $|\vec{v}|>C_R$, even though rest mass cannot be defined for them. we can determine the intrinsic momentum of a tachyon by relation $|\vec{p}_\infty|=\sqrt {\vec{p}^2-\frac{E^2}{C_R^2}}$. As tachyons and constons both no having rest mass, all reference frames are formed by tardyons, so the velocity of transformations between inertial frames are restricted by $|V|<C_R$.\\

(5) The familiar Maxwell equations do not satisfy the principle of relativity in the limit $C_R\rightarrow \infty$, i.e. case of Galilean transformation with $\gamma(V)=1 $. Therefore this trivial case can be excluded from the generalized Lorentz transformations by virtue of Einstein relativity and a finite value for $C_R$. From the well-established electrodynamics, $c$ is the propagating speed of electromagnetic waves or the speed of photons in vacuum, whose value can be measured in experiments. Nevertheless, in conventional relativity, the Relativity Constant is identified with the speed of light by the assumption of the constant speed of light, and is also the same as the constant in electrodynamics $c=C_R$. It is usually believed that the conventional relativity will be challenged if any superluminal phenomenon is observed in real world. However, our formalism does not rely on the assumption of constant speed of light, and treat the Relativity Constant $C_R$ and the speed of light $c$ as conceptually different quantities with principle different values in reality. Given the current experimental constraints, one can take $C_R=c$. Should any superluminal phenomenon from experiments is confirmed, two situations can occur: If the superluminal particle is tachyonic, i.e. it satisfies $E^2-\vec{p}^2c^2<0$ , then the relation $C_R=c$ needs not be corrected. On the other hand, if the superluminal particle satisfies $E^2-\vec{p}^2c^2>0$, then one can conclude that the Relativity Constant is not equal to the speed of light $C_R\neq c$, and the speed of light $c$ is no longer invariant in different inertial frames. The theory of relativity can still work if the value for $C_R$ is corrected. As required by the covariance of physics laws, the Relativity Constant $C_R\neq c$ should replace the speed of light $c$ in the Maxwell equations. Then the photon is a tardyon instead of a conston, with non-zero rest mass (analogous of the story of the neutrino)~\cite{22,23}. As a result, electrodynamics will have to be modified by mass effect of photon~\cite{24,25,26,27,28,29}. In this case, the theory of the electrodynamics will be corrected for both the effect of replacing $C_R$ with $c$ and the mass effect of photon.\\

It is worth pointing out that the existance of constons means that the invariant speed can be reached, then the generalized Lorentz transformation can be derived by studying kinematics. Still, our discussion has shown that Special Relativity can be derived only from tardyons with $|\vec{v}| < C_R$, without the need for any knowledge of the particles with $|\vec{v}| \geq C_R$. Those particles, if they exist at all, are not required {\it a priori} to formulate a consistent relativistic theory. So the existance of constons is sufficient foundation rather than essential foundation for Special Relativity. The assumption of constant speed of light restricts the theory to the possibility that the Relativity Constant $C_R$ is nothing but the speed of light. Nonetheless, people tend to cast doubt on Special Relativity from various discussions on superluminal phenomena~\cite{30,31,32,33,34,35,36,37,38}. From our point of view, the Relativity Constant $C_R$ is connected to the interacting nature of all particles, and should be measured by experiments. Current results are consistent with $C_R=c$ to good precision, so constant speed of light becomes a strong observational fact. However, should superluminal phenomenon ever be observed in the future, two distinct possibilities have to be taken into account: (i) it could be that the superluminal particle has rest mass, then the numerical value for the Relativity Constant $C_R$  must be corrected, or (ii) the superluminal particle has no rest frame, then it should be considered as a candidate for tachyon with a real mass parameter, which might solve the Dark Matter problem. Superluminal phenomenon and relativity will be consistent with each other in either case. Without assuming a constant speed of light, our new derivation has strengthened the foundation of Special Relativity. As a result, Special Relativity should apply to a wider realm of physics regardless of whether or not particles with $|\vec{v}|= C_R$ exist or whether or not superluminal phenomenon is discovered.

\begin{acknowledgments}
Y. D. would like to thank Liang Dai for many useful discussions and help. We thank Xiaotong Song and Jianhui Dai for useful suggestions. We also thank Hanqing Zheng and Xuean Zhao for boosting comments to this work.
\end{acknowledgments}


\end{document}